\documentclass[10pt]{article}

\usepackage{ragged2e}       
\usepackage{soul}           
\usepackage{xcolor}
\usepackage{graphicx}
\usepackage{hyperref}
\usepackage{amsmath}
\usepackage{subcaption}
\usepackage{amssymb}
\usepackage{booktabs}
\usepackage{fullpage}

\definecolor{my_purple}{RGB}{255,0,255}
\graphicspath{{paper_figs/}}

\definecolor{REDCOLOR2}{RGB}{0,0,0}

\title{\textbf{\Large Pumping Patterns and Work Done during Peristalsis in Finite{-length} Elastic Tubes}}

\author{
   \small \bf Shashank Acharya \\
   \small Department of Mechanical Engineering\\
   \small Northwestern University\\
      		\and
   \small \bf Wenjun Kou \\
   \small Division of Gastroenterology and Hepatology\\
   \small Feinberg School of Medicine, Northwestern University,\\
      		\and
   \small \bf Sourav Halder \\
   \small Theoretical and Applied Mechanics Program\\
   \small Northwestern University\\
      		\and
   \small \bf Dustin A. Carlson \\
   \small Division of Gastroenterology and Hepatology\\
   \small Feinberg School of Medicine, Northwestern University\\
      		\and
   \small \bf Peter J. Kahrilas \\
   \small Division of Gastroenterology and Hepatology\\
   \small Feinberg School of Medicine\\
      		\and
   \small \bf John E. Pandolfino \\
   \small Division of Gastroenterology and Hepatology,\\
   \small Feinberg School of Medicine\\
      		\and
   \small \bf Neelesh A. Patankar \\
   \small Department of Mechanical Engineering,\\
   \small Northwestern University,\\
   \small e-mail: n-patankar@northwestern.edu
}

\date{}

\begin{document}

\maketitle

\begin{abstract}
Balloon dilation catheters are often used to quantify the physiological state
of peristaltic activity in tubular organs and comment on their ability to propel
fluid which is important for healthy human function. To fully understand this
system's  behavior, we analyzed the effect of a solitary peristaltic wave on a
fluid-filled elastic tube with closed ends. A reduced order model that
predicts the resulting tube wall deformations, flow velocities and pressure
variations is presented. This simplified model is compared with detailed
fluid-structure 3D immersed boundary simulations of peristaltic pumping in
tube walls made of hyperelastic material.  The major dynamics observed in the
3D simulations were also displayed by our 1D model under laminar flow
conditions. Using the 1D model, several pumping regimes were investigated
and presented in the form of a regime map that summarizes the system's
response for a range of physiological conditions. Finally, the amount of work
done during a peristaltic event in this configuration was defined and
quantified. The variation of elastic energy and work done during pumping was
found to have a unique signature for each regime. An extension of the 1D model
is applied to enhance patient data collected by the device and find the work
done for a typical esophageal peristaltic wave. This detailed characterization
of the system's behavior aids in better interpreting the clinical data
obtained from dilation catheters. Additionally, the pumping capacity of the
esophagus can be quantified for comparative studies between disease groups.
\end{abstract}

Keywords: {esophagus, elastic tube flow, \textcolor{REDCOLOR2}{peristalsis}, reduced-order modeling, fluid-structure interaction, immersed boundary method}

\section{Introduction}

Peristaltic flow through cylinders and other geometries has been studied
extensively since the mid 1960s \cite{Jaffrin1971, Fung1968, Burns1967,
Latham1966PhDThesis}. The investigations have ranged from analyses of
specified wall motion on Newtonian \cite{Takabatake1982, Jaffrin1971},
non-Newtonian \cite{Siddiqui1994, Bohme1983} and particulate fluids
\cite{Mekheimer1998, Hung1976, Chrispell2011} to the effects of a prescribed
forcing on the coupled fluid-structure system \cite{Takagi2011, Carew1997,
Griffiths1987}. Both infinite and finite geometries \cite{Li1993,
Griffiths1987} have been considered and the quantitative effects of the
channel geometry (cylindrical vs. rectangular channels) on pumping
characteristics have been established \cite{Burns1967, Takabatake1988}.
{However, one problem that has received little attention is the effect of
peristalsis on fluid-filled elastic tubes that are closed at both ends.}  In
such a setting, fluid inside the tube can neither enter nor leave the system
for the entire duration of peristalsis.  {This configuration is
commonly found in long, slender balloon catheters used to evaluate mechanical
properties and the contractile response of blood vessels (similar to catheters
used during angioplasty)}. It is also found in functional lumen imaging probe
(FLIP) catheters used to characterize the contractility of anatomical
sphincters and esophageal motility.

\begin{figure*}
    \centering{\fbox{\includegraphics[trim=0 100 50 100,clip,width=\textwidth]{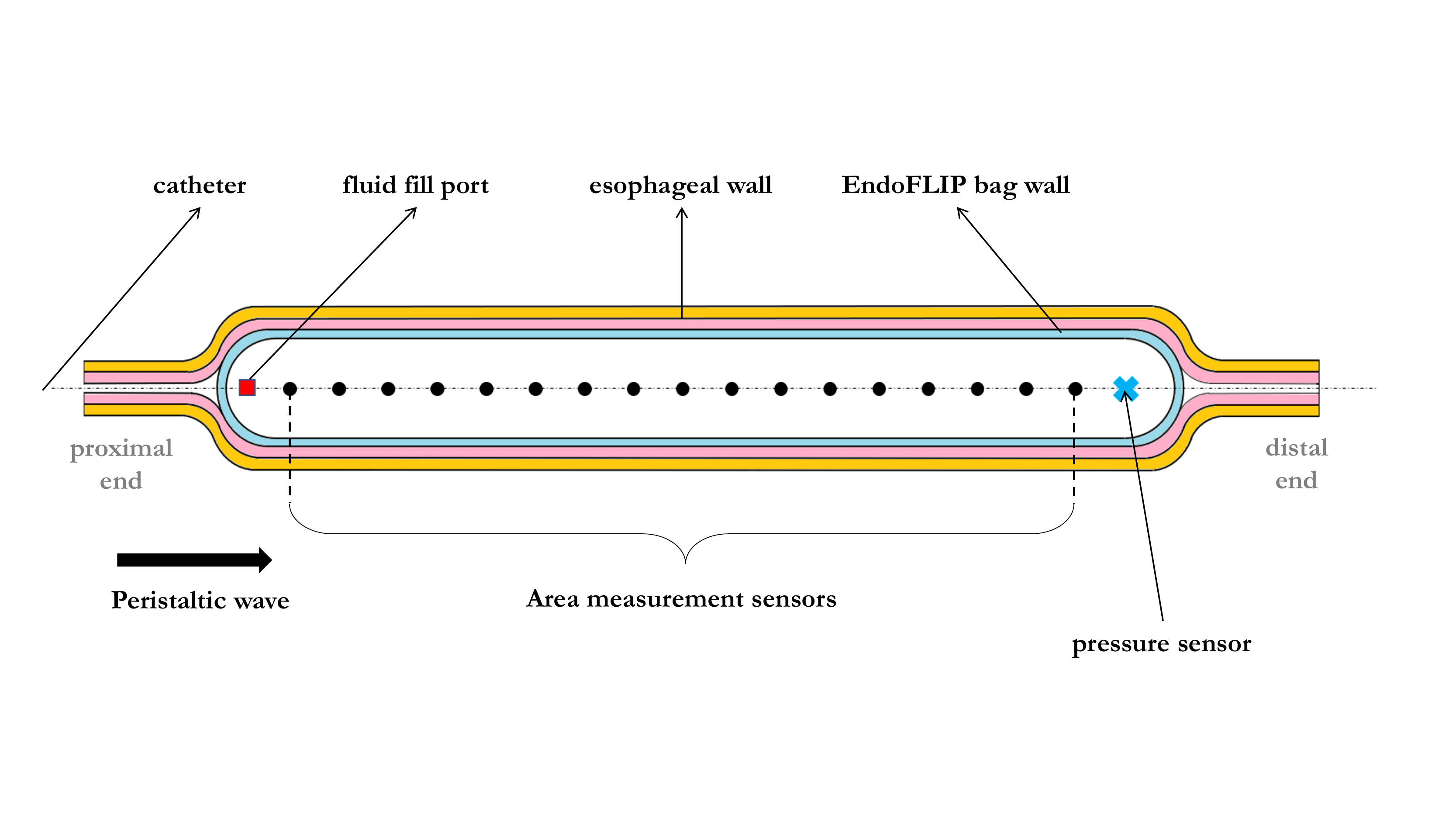}}}
    \caption{Details of the FLIP bag and catheter assembly (positioned within the
    esophagus). The proximal end of the bag points towards the subject's
    throat and the distal end, towards the stomach. \textcolor{REDCOLOR2}{The length of the bag is
        16 cm and diameter at full distension is 22 mm}.}
    \label{fig:EndoFLIP_catheter_closeup}
\end{figure*}

A major assumption utilized to facilitate the mathematical analysis of
peristaltic flow through a tube is that the problem is identical in both the
fixed (lab) frame of reference and in the reference frame attached to the
peristaltic wave. In the latter, the wave is stationary and the shape of the
tube walls does not change with time. When the tube length is finite, this
assumption is not valid \cite{Li1993}. Thus, in order to build a simplified
model for our problem, we turn to the approach taken to
{analyze flow driven by valveless pumping, which occurs in a
finite, periodic domain, and modify the forcing method and boundary
conditions} to reflect the operating conditions for our problem.
Interestingly, the configuration we aim to study first appears in
\cite{Jaffrin1971}, but is subsequently ignored. {To the best
of our knowledge, this configuration does not appear again in any other
studies involving flow in deformable tubes.}

The device we intend to focus our analysis on is the aforementioned FLIP
\cite{Carlson2015}. It consists of a long flexible, hollow catheter, at the
end of which is mounted a polyurethane bag \cite{Regan2012}. The section of
the catheter enclosed by the bag incorporates paired impedance planimetry
sensors that can measure the regional cross-sectional area (CSA) along the
bag's length \textcolor{REDCOLOR2}{as seen in Fig.~\ref{fig:EndoFLIP_catheter_closeup}}. Various
versions of the device exist where the bag length is either 8 or 16 cm; our
focus is on the latter version. Also within the bag at the distal end is a
pressure sensor that captures the fluid pressure at any given time. When
deployed in a patient, the device is positioned such that most of the bag
rests within the esophageal lumen. The bag is then incrementally filled with
saline and the esophageal contractile response is evaluated by monitoring the
internal fluid pressure and CSAs along the length of the bag. When fully
distended with saline, the bag diameter is 22 mm. The external end of the
catheter is connected to a computer that stores the data collected by these
sensors. The bag can be filled or drained by pumping fluid through the
catheter and fluid enters or leaves the bag through a small hole on the
catheter. Figure~\ref{fig:EndoFLIP_catheter_closeup} shows a simplified
representation of the bag-catheter assembly and shows the locations of the
various sensors. For additional details on the device's construction, data
resolution, accuracy and frequency of data collection, see Refs.
\cite{Carlson2015} and \cite{Kwiatek2010}.

\begin{figure*}
  \centering{\fbox{\includegraphics[trim=0 0 0 0,clip,width=\textwidth]{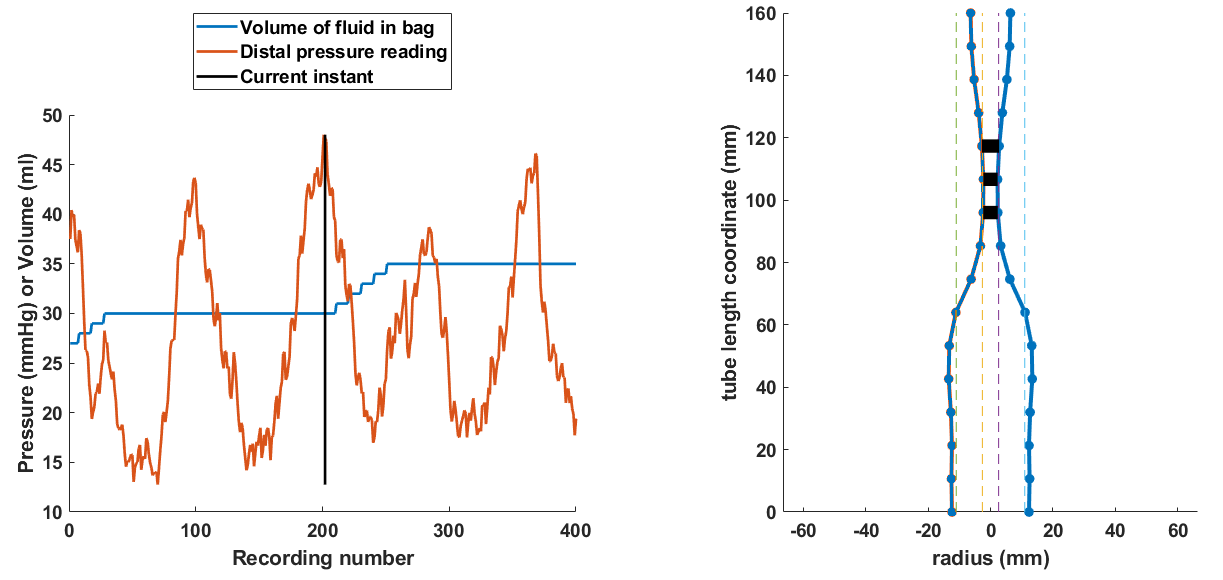}}}
  \caption{Visualization of clinical data collected from the FLIP. Graph on the left
  shows the bag volume and pressure variation over 40 seconds of time (10 Hz
  recording frequency). Figure on the right shows the profile of the
  esophageal lumen at the instant marked by the vertical black bar on the left
  graph (at recording number 200). Dotted lines show the range of measurable
  cross-sectional area (21 - 380 $\mathrm{mm}^2$, 5.2 - 22 mm diameter) \cite{Carlson2015}.}
  \label{fig:FLIP_5MT_example}
\end{figure*}


\subsection{Operating Details and Simplifying Assumptions} \label{opr_details}

Representative FLIP data obtained from a subject can be visualized as shown in
Fig. \ref{fig:FLIP_5MT_example}.  \textcolor{REDCOLOR2}{During examination, the subject is in a
supine position.} The left panel shows the pressure and volume inside the bag
as a function of time. It should be noted that the volume change is controlled
by the physician conducting the procedure and that the pressure change is a
consequence of esophageal contractility in response to distension. The right
panel of the figure shows the axisymmetric profile of the  tube at the instant
in time indicated by the black vertical line on the left panel. The dotted
lines superimposed on the tube profile show the maximum and minimum diameters
that can be accurately measured by the planimetry sensors.  The esophageal
contraction is highlighted with the three black bands where the CSA is the
least. On the pressure curve, we see four peaks indicating that four
peristaltic waves have passed over the bag for the duration of this plot.  The
image of the bag profile is oriented such that peristalsis begins at the top
and travels downward. Measurement of area is not continuous along the bag's
length. Rather, area is measured at 16 locations at 1 cm intervals and
intermediate values are  interpolated to construct the lumen profile
\cite{Jain2019}.

With the device located in the esophagus, the passage of a peristaltic wave
causes the (axisymmetric) profile of the bag and fluid pressure within it to
change. During peristalsis, the esophageal wall may or may not approximate
(i.e. come in contact with) the catheter. It is difficult to predict the
relationship between flow rate and pressure drop across the contraction zone.
The Reynolds number ($Re$) of the system is estimated to be 660 when using the
following values: density $\rho=1000\ \text{kg/m}^{3}$, viscosity $\mu=0.001\
\text{Pa}\cdot\text{s}$, tube diameter $D=22\ \text{mm}$ and peristaltic wave
speed $c=3\ \text{cm/s}$ \textcolor{REDCOLOR2}{\cite{Kou2015ajpgi}}. If the
flow inside is assumed to be similar to pipe flow at every location, then the
flow  is laminar. But the Reynolds number in the contraction zone is difficult
to estimate so the nature of flow at the neck is unknown. As such, we will
analyze two flow types, (corresponding to low and high $Re$): 1) parabolic
flow everywhere and 2) assuming that that a simple friction factor can be used
to relate flow rate and pressure drop in the entire domain.
\textcolor{REDCOLOR2}{As the subject is in a supine position, we assumed that
the effect of gravity on the device is negligible. Depending on the 
positioning of the esophagus in the subject, the catheter can become slightly
curved during measurement. The impedance planimetry sensors measure only local
lumen area. Thus, curvature information is not available during clinical
examination or subsequent analysis. However, fluoroscopic imaging
\cite{Ott1989} shows that the curvature is not significant, thus we assume a
straight geometry to model this problem. During the procedure, bag volume is
changed from 0 to 70 mL in 10 mL steps. The analysis presented in this work
considers peristaltic activity occurring at 30 or 40 mL of distension where
the esophagus is gently stretched and circumferential wall strains are not
high. Due to the inflated nature of the bag, the lumen profile tends to be
largely circular.}

Our goal was to build a simple model to analyze this system and understand the
relationships between the tube profile, internal bag pressure, esophageal wall
stiffness and the intensity of the peristaltic contraction. With sound
mathematical foundations, the work done by the esophageal walls to pump fluid
within the bag can be defined. {This fast, simple model will
eventually be deployed at point-of-care to estimate pressure distributions and
work done in real time.} Armed with this knowledge, we can quantify
peristaltic work of the esophagus, which may prove valuable in the evaluation
of dysphagia.


\section{Mathematical Details of the 1-D Model} \label{oneDmodeldetails}

\textcolor{REDCOLOR2}{Incompressible} flow in a tube with deforming walls can be modeled using the system of
equations given by (\ref{eq:continuity}) and (\ref{eq:momentum}). Here,
$A(x,t)$ is the tube area and $u(x,t)$ is the area-averaged fluid velocity in
the axial direction. The axial coordinate along the tube length is denoted by
$x$ and time by $t$. {Equation (\ref{eq:continuity}) is the
continuity equation that relates local changes in tube area and fluid velocity
to conserve volume. Equation (\ref{eq:momentum}) describes the conservation of
momentum with a general flow resistance term $2\tau_R/(\rho R)$ due to fluid
viscosity and a momentum correction factor of unity \cite{Wang2014}.} These
equations have been widely used to describe valveless pumping
\cite{Manopoulos2006, Bringley2008}, and a detailed derivation of these
equations can be found in \cite{Ottesen2003}.

\begin{equation} \label{eq:continuity}
    \frac{\partial A}{\partial t}+\frac{\partial\left(Au\right)}{\partial x} = 0
\end{equation}

\begin{equation} \label{eq:momentum}
    \frac{\partial u}{\partial t} + u\frac{\partial u}{\partial x} = 
    -\frac{1}{\rho}\frac{\partial p}{\partial x}-\frac{2\tau_{R}}{\rho R}.
\end{equation}

{In the context of peristalsis-driven flow in the esophagus, fluid
velocities are much smaller than the velocity at which a disturbance propagates
in the esophageal wall. Thus, we expect no shocks or discontinuities in the
problem which allows us to use the non-conservative forms of the continuity
and momentum equations. For smooth solutions, both conservative and
non-conservative versions of the governing equations are equivalent and will
lead to the same numerical solution.}

To close the system of equations, we assume that a constitutive law in the
form of an explicit relationship between pressure $p$ and area $A$ exists.
This relationship is commonly referred to as a ``tube law''. For our
application, we assume that the change in pressure is proportional to the
change in area. Experiments carried out with the FLIP show that the distal
esophageal walls follow this behavior \cite{Kwiatek2011} and a similar form of
the tube law is also used in \cite{Bringley2008} and derived in
\cite{Whittaker2010} using shell theory. {The
external pressure is assumed to be zero for all time} and thus the transmural
pressure is equal to the pressure inside the tube. A damping term is
introduced to {regularize} the system of equations. The form of the tube law
used is given by
\textcolor{REDCOLOR2}{
\begin{equation} \label{eq:tubelaw_dim}
    p = K\left(\frac{A}{\theta A_0} - 1\right) + Y\frac{\partial A}{\partial t}.
\end{equation}
}

\noindent Here, $K$ is a measure of the wall stiffness and has the units of
pressure. Its exact value depends on the Young's modulus of the muscle wall
and the ratio of its thickness to the undeformed radius.
\textcolor{REDCOLOR2}{The tube's undeformed, reference area $A_0$, is changed
by an activation factor $\theta=\theta(x,t)$ to introduce the effect of a
contraction}. A damping term with coefficient $Y$ is also introduced. By a
simple manipulation using the continuity equation, we can see that the
introduction of this term leads to a diffusion term in the momentum equation:

\textcolor{REDCOLOR2}{
\begin{equation} \label{eq:tubelaw_nondim_dx}
    p = K\left(\frac{A}{\theta A_0} - 1\right) + Y\frac{\partial A}{\partial t} = 
    K\left(\frac{A}{\theta A_0} - 1\right) - Y\frac{\partial \left(Au\right)}{\partial x}
\end{equation}}

\textcolor{REDCOLOR2}{
\begin{equation}
    \frac{\partial p}{\partial x} = \frac{K}{A_{0}}\frac{\partial }{\partial x} \left(\frac{A}{\theta}\right)
    - Y\frac{\partial^{2}\left(Au\right)}{\partial x^{2}}.
\end{equation}}

Thus, the addition of this term helps stabilize the numerical solution. A
similar manipulation was also used in \cite{Wang2014} to eliminate $p$. {In
our analysis, the damping is kept small so as to ensure that the linear part
of the tube law is the major contributor to fluid pressure for all regimes of
operation.} Note that $A(x,t)$ and $\theta A_0$ are not equal. The former is
an unknown whose value must be found as part of the solution. Before
proceeding, it is important to make note of the viscous shear stress term in
the momentum equation. {Several approximations can be made to write $\tau_R$
as a function of tube area and local fluid velocity. For our analysis, we
choose to study the system when 1) the flow is parabolic everywhere with
viscosity $\mu$ and, 2) a viscous term that uses a constant friction factor
$f$ is used to compute flow stresses. In the former case, the viscous
resistance term $2\tau_{R}/(\rho R)$ is equal to $8\pi\mu u/(\rho A)$ and is
$f u |u|/(2D)$ for the latter.}


\subsection{Non-dimensional Version of the Governing Equations}

The following expressions were chosen for each of the dimensional variables to
obtain non-dimensional versions of the continuity, momentum and the tube law
equations:

\begin{equation}
A=\alpha A_{0}, \qquad t=\tau\frac{L}{c}, \qquad u=Uc,
\end{equation}

\begin{equation}
p=PK \qquad \text{and} \qquad x=\chi L.
\end{equation}

\begin{figure*}
    \centering
    \fbox{\includegraphics[trim=0 190 0 170,clip,width=\textwidth]{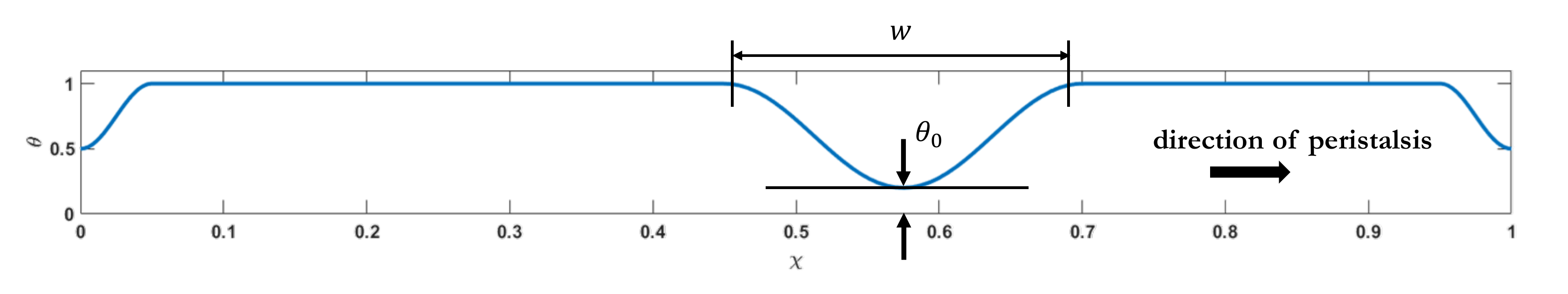}}
    \caption{Visualizing the peristaltic activation wave}
    \label{fig:activation_wave_shape}
\end{figure*}

\noindent Here, speed of the peristaltic wave is denoted by $c$ and $L$ is the
length of the FLIP bag. Substituting these into Eqs. (\ref{eq:continuity}),
(\ref{eq:tubelaw_nondim_dx}) and the parabolic version of Eq.
(\ref{eq:momentum}) gives

\begin{equation} \label{eq:continuity_nondim}
    \frac{\partial\alpha}{\partial\tau}+\frac{\partial\left(\alpha U\right)}{\partial\chi} = 0,
\end{equation}

\begin{equation} \label{eq:momentum_nondim}
    \frac{\partial U}{\partial\tau} + U\frac{\partial U}{\partial\chi} + 
    \psi\frac{\partial P}{\partial\chi}
    + \beta\frac{U}{\alpha} = 0, \qquad \text{and}
\end{equation}

\begin{equation} \label{eq:tubelaw_nondim}
    \textcolor{REDCOLOR2}{P=\left( \frac{\alpha}{\theta}-1 \right) - \eta\frac{\partial\left(\alpha U \right)}{\partial\chi}},
\end{equation}

\noindent where the following non-dimensional numbers $\psi=K/(\rho c^2)$,
$\eta=(YcA_0)/(KL)$ and $\beta = 8\pi\mu L/(\rho A_0 c)$  emerge. Here, $\eta$
is a nondimensional tube damping parameter and $\beta$ is a flow resistance
coefficient. Eventually, the non-dimensional pressure $P$ in the momentum
equation (\ref{eq:momentum_nondim}) will be replaced with the right hand side
of Eq. (\ref{eq:tubelaw_nondim}). Thus, the total number of coupled equations
to be solved is two with $\alpha$ and $U$ as the unknowns.
{When working with the friction factor version of the momentum
equation, the non-dimensional number for flow resistance $\beta$ becomes
$\left(fL/4\right)\sqrt{\pi/A_0}$ and $\psi$ remains unchanged and the viscous
resistance term in the momentum equation takes the form $\beta
U|U|/\sqrt\alpha$.}


\subsection{Boundary Conditions and Peristaltic Wave Input}

It is clear that if $\eta$ is exactly zero, Eqs.~(\ref{eq:continuity_nondim}),
(\ref{eq:momentum_nondim}) and (\ref{eq:tubelaw_nondim}) allow only one boundary
condition each for $\alpha$ and $U$ as the system is hyperbolic. {However,
during normal operation, the fluid volume inside the bag is constant because
saline pumping is halted and the bag ends are sealed. Thus, fluid velocity at
both ends must be set to zero which is not admissible with $\eta=0$. In
addition to this, hyperbolic systems often present convergence issues. To
address these issues, we set $\eta\neq0$ and add a smoothing term
$\epsilon\alpha_{\chi\chi}$ to the continuity equation} \cite{LeVeque1990}.
Following this step, substituting Eq.~(\ref{eq:tubelaw_nondim}) into
Eq.~(\ref{eq:momentum_nondim}) leads to second derivatives for $\alpha$ and $U$
requiring two BCs each for $\alpha$ and $U$ at both ends. The bag is attached
to the catheter at both ends and tapers off to a point. Thus, boundary
conditions for velocity at the ends,

\begin{equation} \label{eq:velBC}
    U\left(\chi=0,\tau\right)=0\qquad\text{and}\qquad U\left(\chi=1,\tau\right)=0
\end{equation}

\noindent are straightforward but the boundary conditions for nondimensional area
$\alpha$ are unclear. {Fixing the area to some nonzero constant
value is equivalent to fixing the pressure and leads to an inconsistent
problem definition. Boundary conditions for $\alpha$ must be consistent with
BCs for $U$. We arrive at these by turning to} Eq.~(\ref{eq:momentum_nondim})
with $U=0$ at the ends which results in $\partial P/\partial \chi = 0$. From
Eq.~(\ref{eq:tubelaw_nondim}), we see that this leads to

\begin{equation}
\frac{\partial \alpha}{\partial \chi} - \eta \left( U\frac{\partial^2\alpha}{\partial\chi^2} + 
        2\frac{\partial\alpha}{\partial\chi}\frac{\partial U}{\partial\chi} + \alpha\frac{\partial^2 U}{\partial\chi^2} \right) = 0.
\end{equation}

\noindent In the limit of $\eta\to 0$, this is equivalent to requiring
$\partial\alpha/\partial\chi=0$ at both boundaries. Thus, we choose

\begin{equation} \label{eq:areaBC}
    \left.\frac{\partial \alpha}{\partial \chi}\right|_{\chi=0,\tau}=0\qquad\mathrm{and}\qquad
    \left.\frac{\partial\alpha}{\partial\chi}\right|_{\chi=1,\tau}=0
\end{equation}
as Neumann-type boundary conditions for $\alpha$ at the tube ends.
{For nonzero damping, this BC is equivalent to changing the viscous
resistance by an amount $-\eta\alpha U_{\chi\chi}$ at the ends. This
additional resistance at the boundary will have negligible effect for small
$\eta$. With the additional regularizing terms and applied BCs, the solution
differs from the hyperbolic version at a thin region near the boundary. In the
next section, using the Method of Manufactured Solutions, we show that the
effect of damping on the overall solution is negligible.} It should be noted
that more complicated forms of the tube law can be used which account for
longitudinal curvature, bending and tension \cite{Grotberg2004,
Mcclurken1981} in the tube wall. Such a tube law can include a double
derivative, $A_{xx}$ or a fourth order derivative term $A_{xxxx}$ of the tube
area.  Under this setting, additional boundary conditions for area and
pressure can be applied in a straightforward manner without leading to
inconsistencies in the problem definition. The inclusion of these terms will
lead to higher order spatial derivatives into the governing equations.
{However, in our current analysis, we choose to work with the simplest version
of the tube law as the esophagus' material properties and behavior for these
deformation modes is unknown.}

The last ingredient required to complete the model is specifying an activation
input to mimic peristaltic contraction. We wish to induce a reduction in the
tube area at some location $\chi$ and at some time $\tau$ in a manner that
resembles a traveling peristaltic wave.  The most common approach to induce
contractions is seen in the aforementioned valveless pumping models where an
external activation pressure at a specific location is varied sinusoidally
with time \cite{Bringley2008, Manopoulos2006, Ottesen2003}. This is an
appropriate method of activation for valveless pumping scenarios due to the
fact that the extra pressure is generated due to respiration.
{However, in the esophagus, contractions in area are generated
due to a contraction of muscle fibers within the esophageal wall. As such, we
add an activation term $\theta$ to the tube law which changes the reference
area of the tube when activated.} The tube law used is

\begin{equation} \label{eq:activated_tube_law}
    P=\left(\frac{\alpha}{\theta}-1\right)-\eta\frac{\partial\left(\alpha U\right)}{\partial\chi},
\end{equation}

\noindent with the activation term given by

\begin{equation}
\theta(\chi,\tau)=
  \begin{cases}
  1 - \left(\frac{1-\theta_{0}}{2}\right)  \Big[1 \ +   \\ 
  \qquad \sin\left(\frac{2\pi}{w}\left(\chi-\tau\right) +
        \frac{3\pi}{2}\right)\Big], & \tau-w\leq\chi\leq\tau \\
  1, & \text{otherwise.}
\end{cases}
\end{equation}

\noindent The non-dimensional width of the peristaltic wave is denoted by $w =
W/L$ with $W$ denoting the dimensional contraction width. {As mentioned
earlier, the CSA at the tube ends in the device tapers off to a point on the
catheter, so a small correction is made to the activation term to account for
the reduced CSA at the ends.} The final form of the activation wave is plotted
in Fig. \ref{fig:activation_wave_shape}. The value of $\theta_0$ specified
represents the reference area of the tube at the strongest part of the
contraction. Physiologically speaking, $\theta_0$ is a parameter that
determines the strength of the contraction. It represents the ratio by which
the zero-stress luminal area is reduced (i.e. the reference luminal area
$A_0$) due to muscular contraction. The smaller the value of $\theta_0$, the
smaller is the activated reference area and the greater is the contraction intensity.
When $\theta_0=1$, there is no contraction and the system is fully at rest.
Since the velocity scale was chosen to be the same as the speed of the
peristaltic wave, the non-dimensional wave speed is 1.0.


\begin{figure}
  \centering{\includegraphics[trim=100 270 130 360,clip,width=0.5\linewidth]{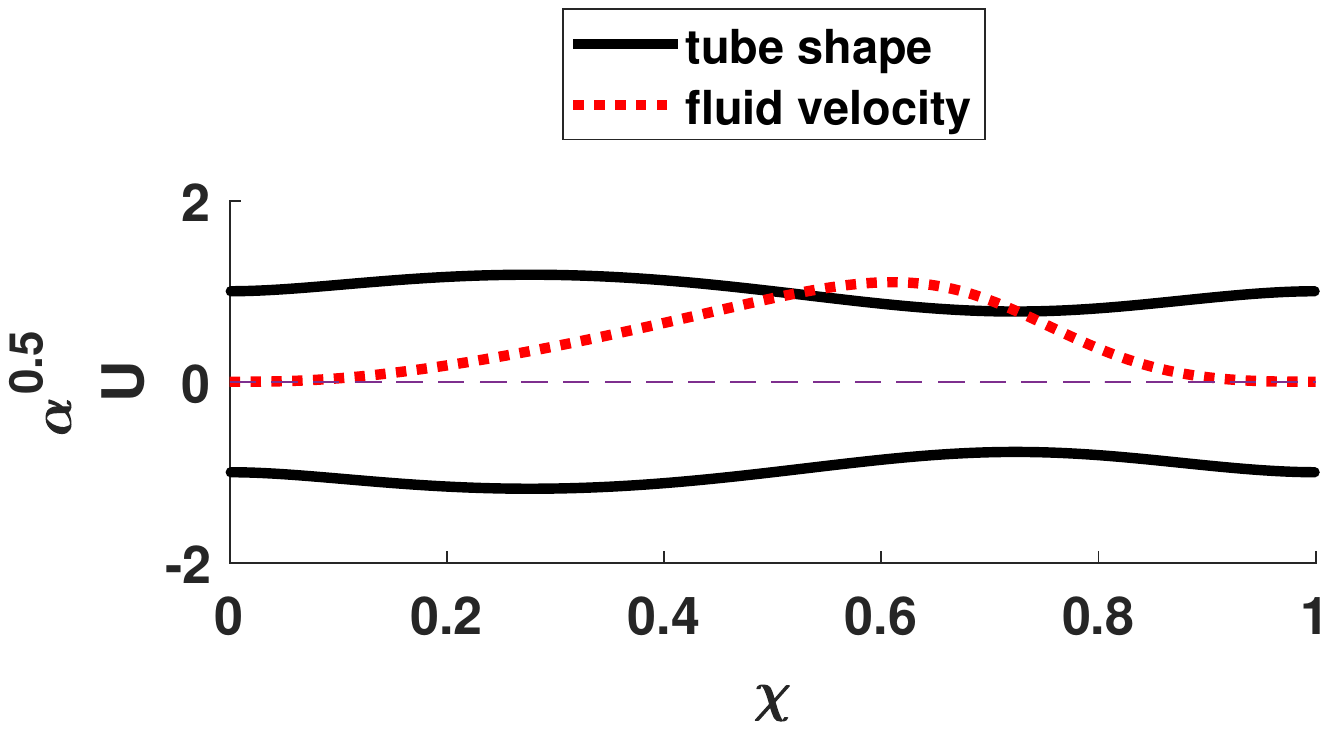}}
  \caption{Tube shape (black lines) and fluid velocity (dotted red line) corresponding to the manufactured solution
  given by Eqs. (\ref{eq:mms_area}) and (\ref{eq:mms_vel}) for $\tau=0.38$}
  \label{fig:mms_chosen_soln}
\end{figure}

\subsection{Numerical Solution of the System of Equations}

The non-dimensional pressure term $P$ in the momentum equation is eliminated
using Eq. (\ref{eq:activated_tube_law}), which leads to the system of equations

\begin{equation}
\frac{\partial\alpha}{\partial\tau} + \frac{\partial(\alpha U)}{\partial\chi} = \epsilon\frac{\partial^2\alpha}{\partial\chi^2},
\end{equation}

\begin{equation}
\frac{\partial U}{\partial\tau} + U\frac{\partial U}{\partial\chi} + \beta\frac{U}{\alpha} + 
        \psi\frac{\partial}{\partial\chi}\left(\frac{\alpha}{\theta}\right) = 
        \zeta\frac{\partial^2}{\partial\chi^2}\left(\alpha U\right),
\end{equation}

\noindent with boundary conditions given by Eqs.~(\ref{eq:velBC}) and
(\ref{eq:areaBC}). The product of $\eta$ and $\psi$ is denoted by $\zeta$. As
explained earlier, a smoothing term $\epsilon\left(\alpha_{xx}\right)$ was
added to the right hand side of Eq.~(\ref{eq:continuity_nondim}) to accelerate
convergence and reduce computation time. We set zero velocity initial
conditions and specify a spatially-constant initial condition for area that
depends on the volume of fluid present in the bag,

\begin{equation} \label{eq:velareaIC}
    U\left(\chi,\tau=0\right)=0  \qquad\qquad  \alpha\left(\chi,\tau=0\right)=\alpha_{\text{IC}}.
\end{equation}

\noindent This set of equations is solved using the \textit{pdepe} routine in
MATLAB. For the sake of clarity, we rewrite the final version of the governing
equations and the boundary conditions in the form solved using MATLAB below
\textcolor{REDCOLOR2}{(subscripts denote partial derivatives with respect to that variable)}:

\begin{equation} \label{eq:pdes_matlab}
\textcolor{REDCOLOR2}{
    \left[\begin{array}{c}
\alpha_{\tau}\\
U_{\tau}
\end{array}\right]=\left[\begin{array}{c}
\epsilon\alpha_{\chi}\\
\zeta\left(\alpha U\right)_{\chi}
\end{array}\right]_\chi-\left[\begin{array}{c}
U\alpha_\chi+\alpha U_\chi\\
U U_\chi+\beta\frac{U}{\alpha}+\frac{\psi}{\theta^2}\left(\theta \alpha_\chi - \alpha \theta_\chi\right)
\end{array}\right]
}
\end{equation}

\begin{equation} \label{eq:bcs_matlab}
    \left[\begin{array}{c}
0\\
U
\end{array}\right]+\left[\begin{array}{c}
1\\
0
\end{array}\right]\left[\begin{array}{c}
\epsilon\alpha_{\chi}\\
\zeta\left(\alpha U\right)_{\chi}
\end{array}\right]=\left[\begin{array}{c}
0\\
0
\end{array}\right] \quad \text{for}\quad \chi=0,1
\end{equation}

Before proceeding with our analysis of the FLIP device with this mathematical
model, we ensured that the damping terms introduced in the continuity and
momentum equation did not adversely affect the computed solution. The Method of
Manufactured Solutions was used to test the solution approach to ensure that
the system of equations was being solved correctly. A sinusoidally expanding
and contracting tube shape was chosen that generated an oscillating velocity
field within the tube. The solution chosen for $\alpha$ at time $\tau=0$ was

\begin{equation}
\alpha(\chi,\tau=0) = f(\chi) = \frac{256}{9}\left(2\chi^5 - 5\chi^4 + 4\chi^3 - \chi^2\right) + 1
\end{equation}

\noindent and the shape of the tube at all other times was set to vary as a standing
wave and has the form

\begin{equation} \label{eq:mms_area}
\alpha(\chi,\tau) = \left[f(\chi) - 1\right]\cos(\omega \tau) + 1.
\end{equation}

Using the original form of the continuity equation given by
Eq.~(\ref{eq:continuity_nondim}), the velocity field corresponding to the
chosen tube shape was found to be

\begin{equation} \label{eq:mms_vel}
U = \frac{\omega \sin(\omega\tau)}{(f-1)\cos(\omega\tau) + 1}\int(f-1)d\chi.
\end{equation}

A snapshot of the manufactured tube shape and fluid velocity is shown in
Fig.~\ref{fig:mms_chosen_soln}. The left segment of the tube has reached the
maximum area of $\alpha=1.5$ at $\chi=0.25$. At this instant, the right
segment of the tube is beginning to expand and the left segment is contracting
causing fluid to flow into the expanding section of the tube. This corresponds
to a positive fluid velocity as shown by the dotted red curve. Eventually the
right segment of the tube reaches its maximum area and begins to contract and
the cycle continues with fluid traveling from one segment of the tube to the
other in a periodic manner.

\begin{figure}
  \centering{\includegraphics[trim=95 220 100 240,clip,width=0.5\linewidth]{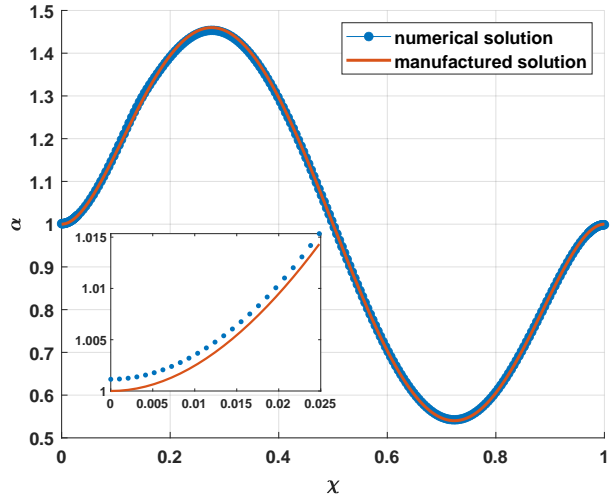}}
  \caption{Comparison of the manufactured solution area and area obtained
  from the numerical solution of Eqs. (\ref{eq:pdes_matlab}) and
  (\ref{eq:bcs_matlab}) for $\tau=3.5$. Inset shows the variation of area
  at the boundary where the gradient was set to zero but the value was
  not kept fixed.}
  \label{fig:mms_compare_area}
\end{figure}

\begin{figure}
  \centering{\includegraphics[trim=95 220 100 240,clip,width=0.5\linewidth]{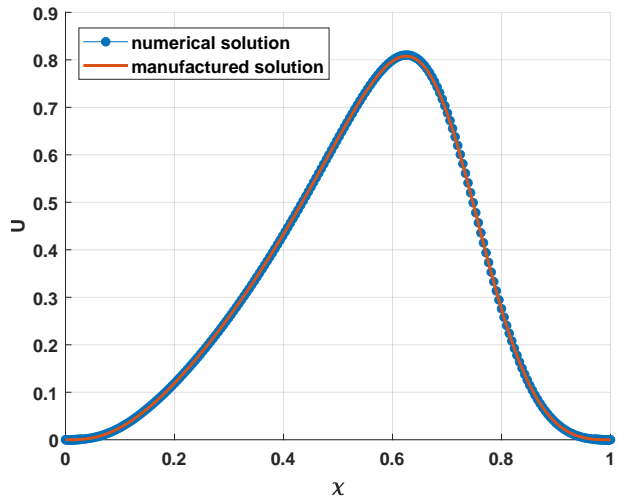}}
  \caption{Comparison of fluid velocity obtained from the numerical solution of the
  regularized equations to the manufactured solution given by Eq.~(\ref{eq:mms_vel}) at $\tau=3.5$}
  \label{fig:mms_compare_vel}
\end{figure}

Using the nondimensional form of the momentum equation
(\ref{eq:momentum_nondim}), the source term corresponding to the chosen
solutions was found with values of $\beta$ and $\psi$ set to $10^3$ and
$\omega=10$. Using the regularized version of the governing equations given by
Eqs. (\ref{eq:pdes_matlab}) and (\ref{eq:bcs_matlab}), along with $\zeta= 0.01$
and $\epsilon= 10^{-4}$, the system was then solved to obtain $\alpha$ and
$U$. Relative tolerance for the solver was set to $10^{-7}$. A comparison of
the manufactured solution to the computed solution is given in Figs.
\ref{fig:mms_compare_area} and \ref{fig:mms_compare_vel}. The agreement
between the chosen and numerically computed solutions was found to be
satisfactory indicating that the equations were being solved correctly and
that the effect of the regularizing terms on the overall solution was
negligible.

\section{Dynamics Displayed by the 1-D Model} \label{dynamics_1d}

\subsection{Comparison with 3D Immersed Boundary Simulations}

Before proceeding with a discussion on the regimes and computing peristaltic
work, we set out to further validate our 1D model with detailed
fluid-structure interaction simulations using the immersed boundary (IB)
method. {It has already been stated that the nature of flow
within the device is not definitively known to be either laminar or fully
developed turbulent flow. As explained in Sec.~\ref{opr_details}, the Reynolds
number for flow in the tube is estimated to be around 660, indicating that the
flow is likely to be laminar.} {For healthy individuals, the
value of $\psi$ is of the order 100 and $\beta$ is of the order 1000 based on
known values of peristaltic wave speed \cite{Pouderoux1997}, tube stiffness
coefficient $K$ \cite{Kwiatek2011}, and material properties of water. The
material parameters for the 3D IB simulation were chosen such that they were
equivalent to $\psi=100$, $\beta=621$ and $\theta_0=0.05$ and flow was
observed to have a parabolic profile everywhere within the tube.} Following
peristaltic activation, the resulting tube shapes and fluid velocity
variations obtained from both models were then compared.

\begin{figure*}
    \centering
    \begin{subfigure}[b]{.49\textwidth}
    \centering
    \includegraphics[width=0.93\linewidth]{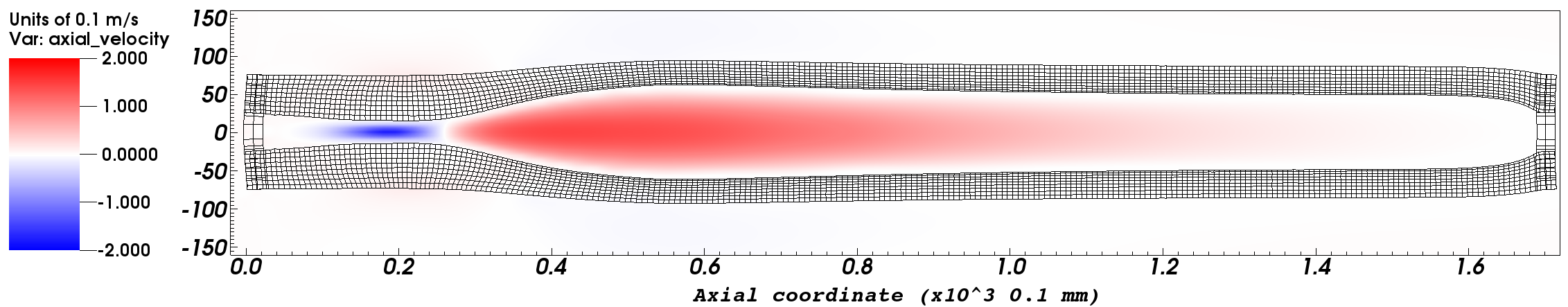}\quad
    \includegraphics[trim=-15 360 40 355,clip,width=1\textwidth]{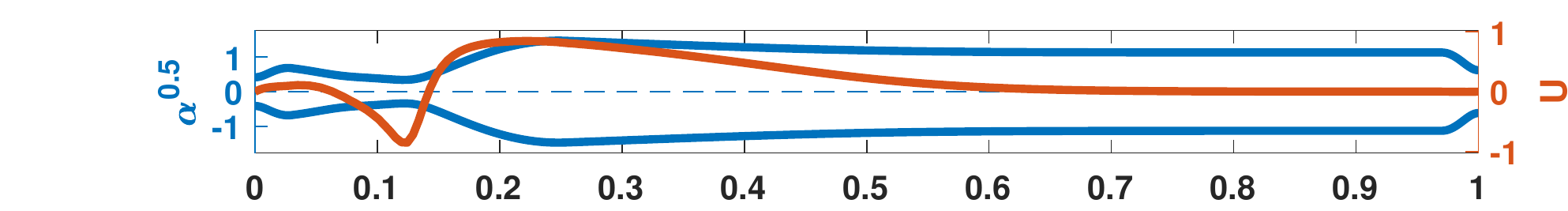}
    \subcaption{Instant 1}
    \label{fig:regime2_inst1}
    \end{subfigure}
    \hfill
    \begin{subfigure}[b]{.49\textwidth}
    \centering
    \includegraphics[width=0.93\linewidth]{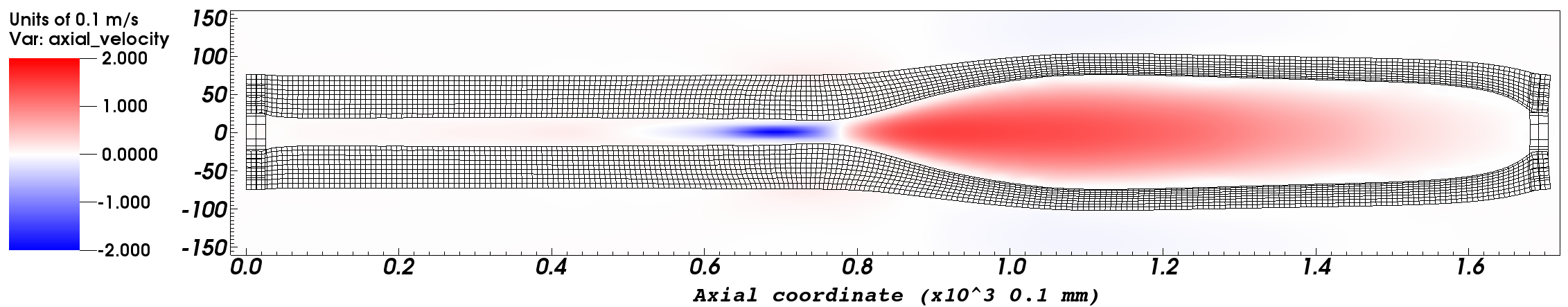}\quad
    \includegraphics[trim=-15 360 40 355,clip,width=1\textwidth]{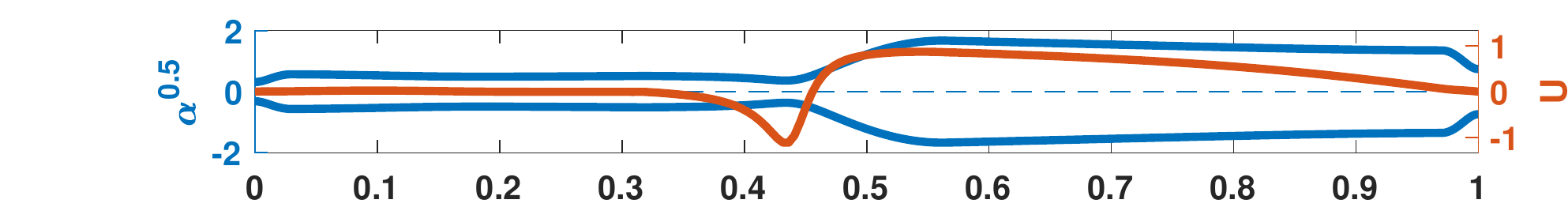}
    \subcaption{Instant 2}
    \label{fig:regime2_inst2}
    \end{subfigure}
    \vskip\baselineskip 
    \begin{subfigure}[b]{.49\textwidth}
    \centering
    \includegraphics[width=0.93\linewidth]{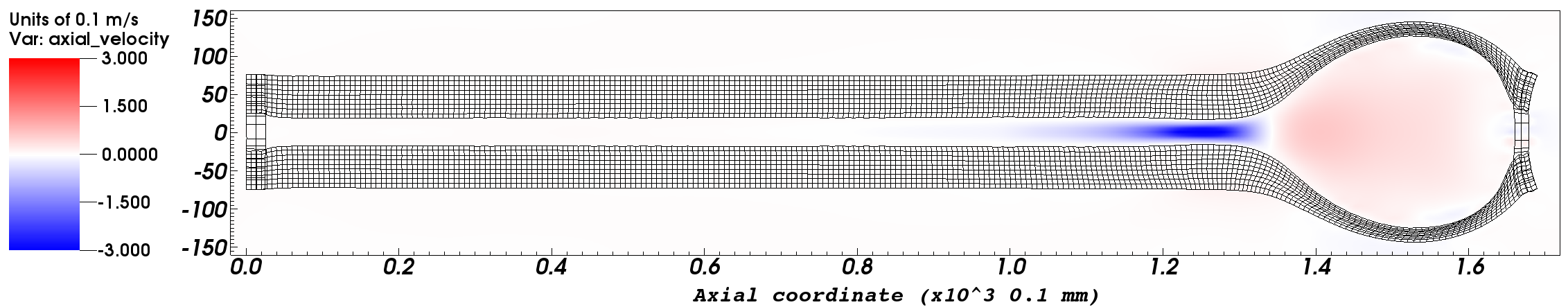}\quad
    \includegraphics[trim=-15 360 40 355,clip,width=1\textwidth]{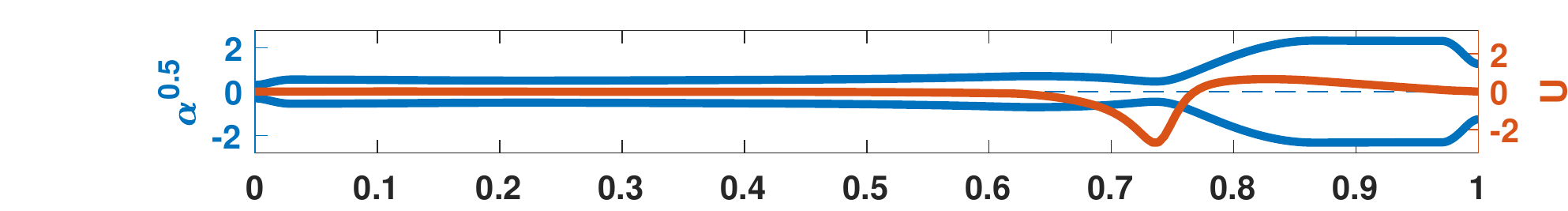}
    \subcaption{Instant 3}
    \label{fig:regime2_inst3}
    \end{subfigure}
    \hfill
    \begin{subfigure}[b]{.49\textwidth}
    \centering
    \includegraphics[width=0.93\linewidth]{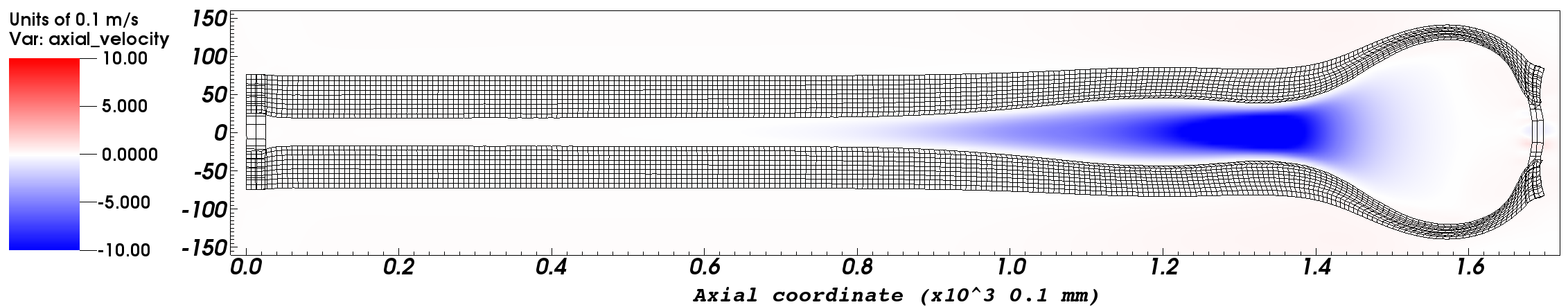}\quad
    \includegraphics[trim=-15 360 40 355,clip,width=1\textwidth]{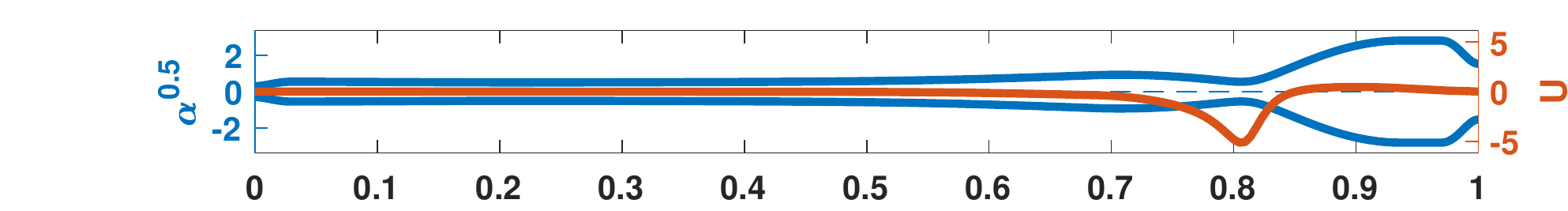}
    \subcaption{Instant 4}
    \label{fig:regime2_inst4}
    \end{subfigure}
    \vskip\baselineskip 
    \begin{subfigure}[b]{.49\textwidth}
    \centering
    \includegraphics[width=0.93\linewidth]{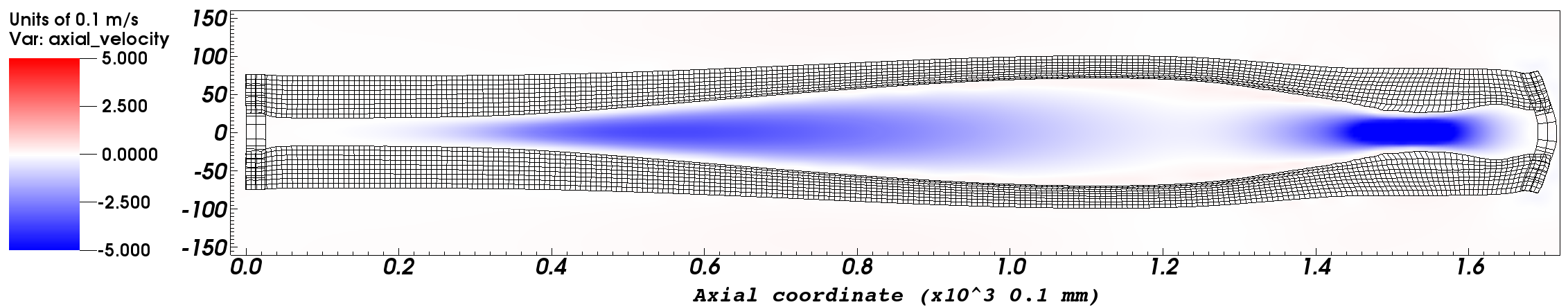}\quad
    \includegraphics[trim=-15 325 40 355,clip,width=1\textwidth]{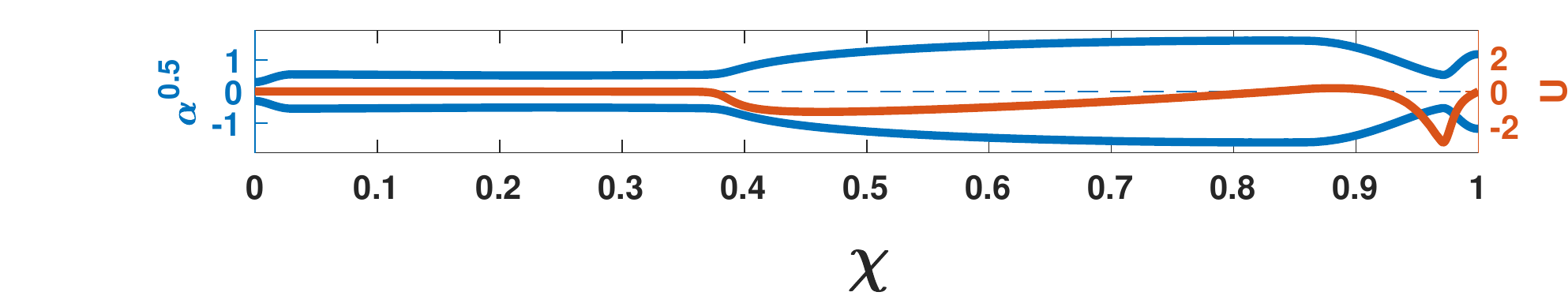}
    \subcaption{Instant 5}
    \label{fig:regime2_inst5}
    \end{subfigure}
    \hfill
    \begin{subfigure}[b]{.49\textwidth}
    \centering
    \includegraphics[width=0.93\linewidth]{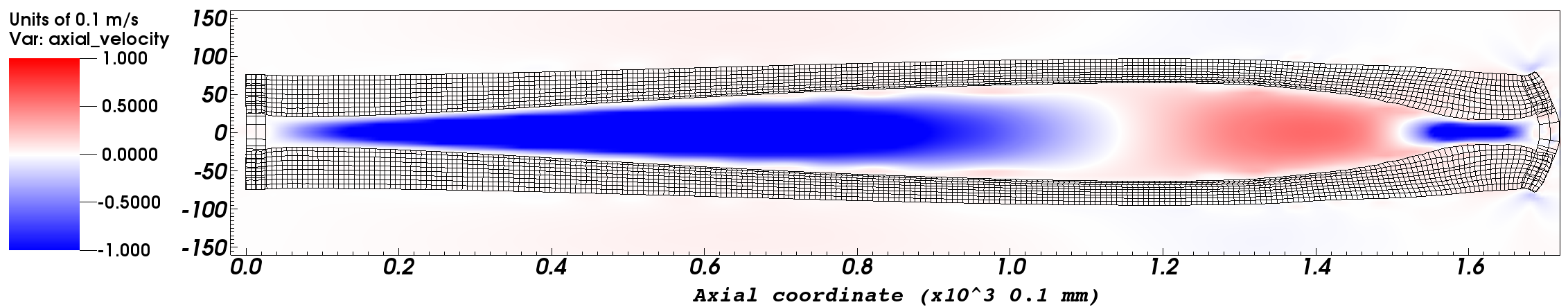}\quad
    \includegraphics[trim=-15 325 40 355,clip,width=1\textwidth]{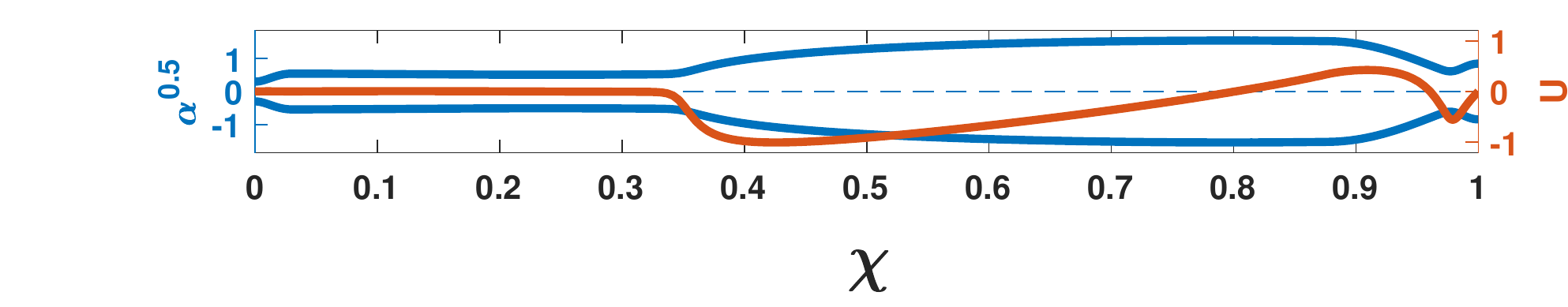}
    \subcaption{Instant 6}
    \label{fig:regime2_inst6}
    \end{subfigure}
    \caption{Comparing tube shapes and velocity fields computed from the 1D model with
    results from equivalent 3D immersed boundary simulations. At each instant,
    the top figure shows the tube mesh along with the internal velocity fields
    along an axial slice obtained from the IB simulations and the bottom
    figure shows the corresponding variations predicted by the 1D model. In
    the latter, the blue curves plot the top and bottom profile of the tube
    using the nondimensional radius $\sqrt{\alpha}$ and the red curve shows
    the variation of velocity $U$ in the tube at each instant as a function of
    $\chi$. The dotted line represents the tube's axis. Values corresponding
    to $\sqrt{\alpha}$ are on the left axis and values for $U$ are on the
    right axis. Negative values of $U$ indicate fluid is traveling in the left
    direction. Contraction wave began at $\tau=0$ and leaves the domain at
    $\tau=1$. Instants 5 and 6 correspond to $\tau=1.10$ and $\tau=1.13$,
    respectively. Figure~\ref{fig:comparison_plot_1d_3d_instant2} compares
    areas and the area-averaged axial velocity with $U$ from the 1D model for Instant 2.}
    \label{fig:comparing_1d_3d}
\end{figure*}

\begin{figure}
\centering\includegraphics[trim=120 235 120 240,clip,width=0.5\linewidth]{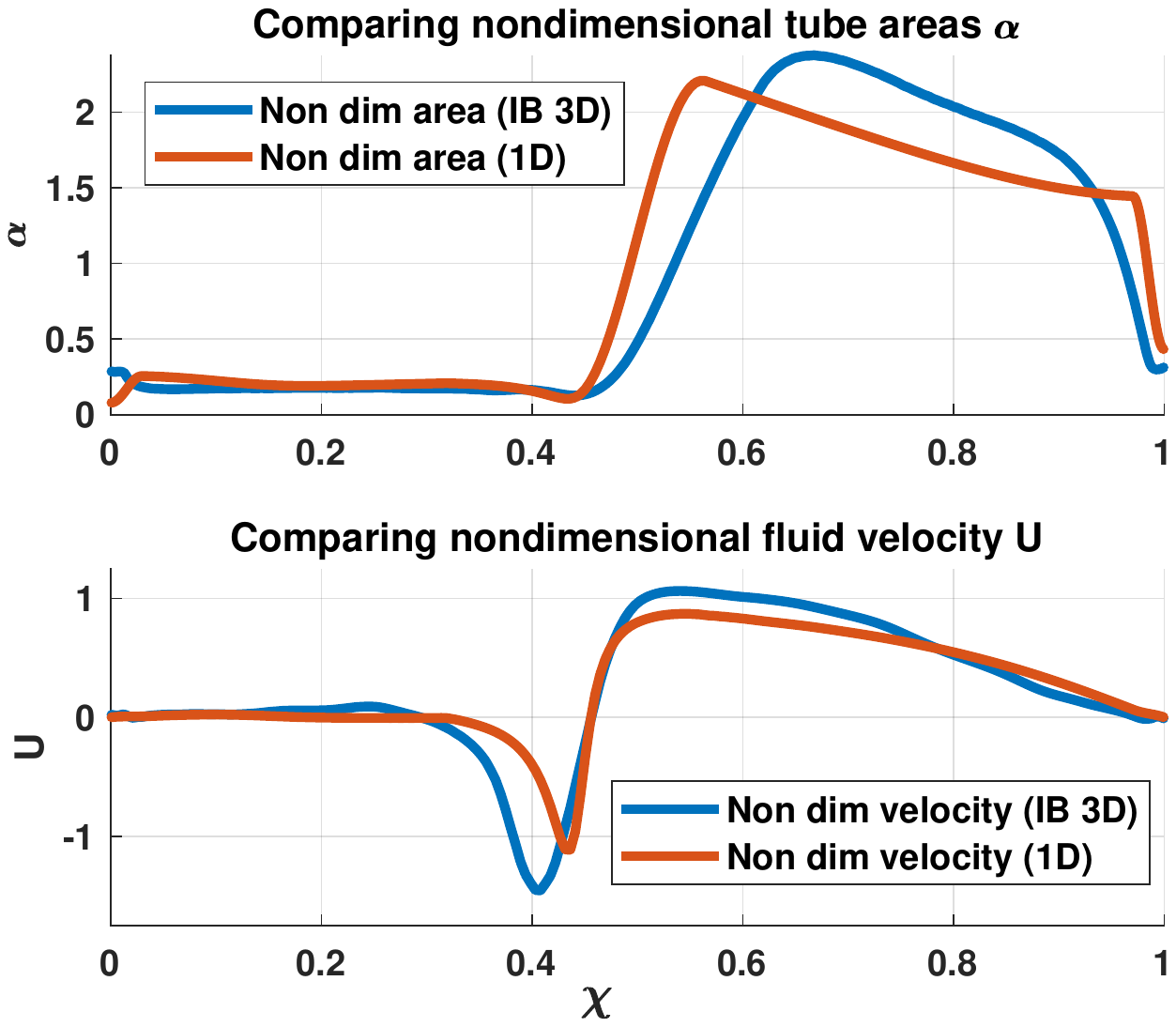}
\caption{Quantitative comparison of nondimensional tube area and fluid velocity
obtained from 3D IB simulations with the numerical solution of the 1D system of
equations. This snapshot corresponds to Instant 2, shown in
Fig.~\ref{fig:comparing_1d_3d} where the contraction is approximately at
$\chi=0.45$.}
\label{fig:comparison_plot_1d_3d_instant2}
\end{figure}

The geometry of the structure used in the IB simulations consists of a long
cylindrical tube with open ends. The two ends are closed using `caps' that are
simply short flat cylinders of specified thickness. The entire structure is
then meshed using hexahedral elements to take advantage of the Adaptive
Anisotropic Quadrature that was developed by Kou \textit{et al.}, in
\cite{Kou2017} and implemented in the open-source immersed boundary framework
IBAMR \cite{IBAMR_cite}. The structure is represented using finite elements
and the combined fluid-structure equations are solved on a Cartesian grid
using the Finite Volume Method. Mathematical details on the interaction
equations, discretization of the Lagrangian and Eulerian domains and temporal
schemes for solving the coupled system can be found in \cite{EGriffith2017}.
Computations were performed on the Northwestern University `Quest' cluster and
on SDSC Comet. The mechanical properties of the tube consists of two
materials. Just like the esophagus, the structure in our IB simulations
consists of fibers embedded in an isotropic matrix material. The matrix
component is represented by a simple Neo-Hookean strain energy function and
the fiber's effect is implemented using a bilinear strain energy function that
computes the stress based on strain in the circumferential direction.
{The combined effect of these layers leads to an approximately
linear relationship between pressure and area corresponding to the tube law
used in the 1D model.} {Additional details on the material properties and
strain energy functions corresponding to each layer can be found in Refs.}
\cite{Kou2017, Kou2015, Kou2015ajpgi}. Similar to the activation method used
in Ref. \cite{Kou2017}, the peristaltic wave is applied as a controlled
reduction in the fiber rest length along the tube. This reduction leads to a
greater generation of circumferential stresses in the tube at the activation
location, the consequence of which is a reduction of the tube area resembling
peristaltic contraction.

The results of our simulations are summarized in
Figs.~\ref{fig:comparing_1d_3d} and \ref{fig:comparison_plot_1d_3d_instant2}.
The figures show the resultant tube shapes generated by the peristaltic
contraction wave as it travels from left to right along the tube length for
both models. As the contraction moves forward, it displaces fluid, causing an
increase in tube area to accommodate the excess fluid. When the pressure in the
segment ahead of the contraction becomes high enough, the contraction opens
up, allowing fluid accumulated in this segment to flow in the opposite
direction and refill the segment of the tube behind the contraction. Figures
\ref{fig:regime2_inst5} and \ref{fig:regime2_inst6} show the tube profile and
fluid velocity after the wave has passed. {At each instant, we see
satisfactory agreement between the velocity variations observed in the 3D
immersed boundary simulations and the non-dimensional velocity profiles
predicted by the 1D model. The tube shapes are also observed to match well
with the shape of the structure computed by the 3D simulations during and
after the activation wave has traveled over the tube's length.}

Encouraged by this agreement between the 1D model and the 3D immersed boundary
simulations, we continued the analysis of the system by probing the 1D model
for a wide range of operating conditions. {Although the
agreement was found only for parabolic flow, we also investigated the system's
response when the viscous term is friction-factor based.} In the next section,
we present the results for both of these scenarios in a ``unified'' regime
map. One of the primary goals for this comparison with the 3D IB simulations
was to check if the introduction of the tube damping and smoothing
terms significantly affected the response of the system. Figures
\ref{fig:comparing_1d_3d} and \ref{fig:comparison_plot_1d_3d_instant2} show
that that the introduction of the regularizing terms not only leaves the
physics of the system unharmed but also aids in stabilizing the numerical
solution which directly leads to more rapid solution times.

{It should be noted that the 3D immersed boundary simulations did not show any
modes of asymmetric collapse or solid-solid contact between the tube walls for
the operating parameters that were tested. The cross section of the tube
remained circular across its entire length for the entire duration of the run.
{The speed at which a disturbance propagates in the system is given by
$\partial p/\partial A\sqrt{A/\rho}$} \cite{Grotberg2004} {and is
approximately 1 m/s for the esophagus. The peristaltic wave speed and internal
fluid velocities are of the order of 1 cm/s indicating that the flow speeds
are always well below the critical value needed for collapse or choking.} At
the location of the contraction, the tube is not passive and the there is no
collapse of the structure. The reduction in area is solely due to active
contraction brought upon by a controlled reduction of the rest length of the
fibers in the tube's material model. It should also be emphasized that the
coupled fluid-structure simulations we have developed are perfectly capable of
capturing any event of collapse if there was a possibility for such an event
to occur. However, for the range of parameters that are relevant to the
operating conditions of this device, collapse is improbable due to the tube's
geometry and the significant amount of fluid in the bag (30 mL or greater).}
\textcolor{REDCOLOR2}{Another thing to note is that as the FLIP only measures area, comparison of
simulation results with realistic three-dimensional structures is not possible
at the moment. In the future, we aim to obtain detailed geometric information
about the esophageal structure from fluoroscopy and 4D-MRI imaging
\cite{Markl4dmri} to compare numerical results with experimentally measured
values.}


\subsection{Types of tube deformations observed during peristalsis}

\begin{figure*}

    \centering
    \begin{subfigure}[b]{0.24\textwidth}
        \centering
        \includegraphics[trim=75 210 60 210,clip,width=\textwidth]{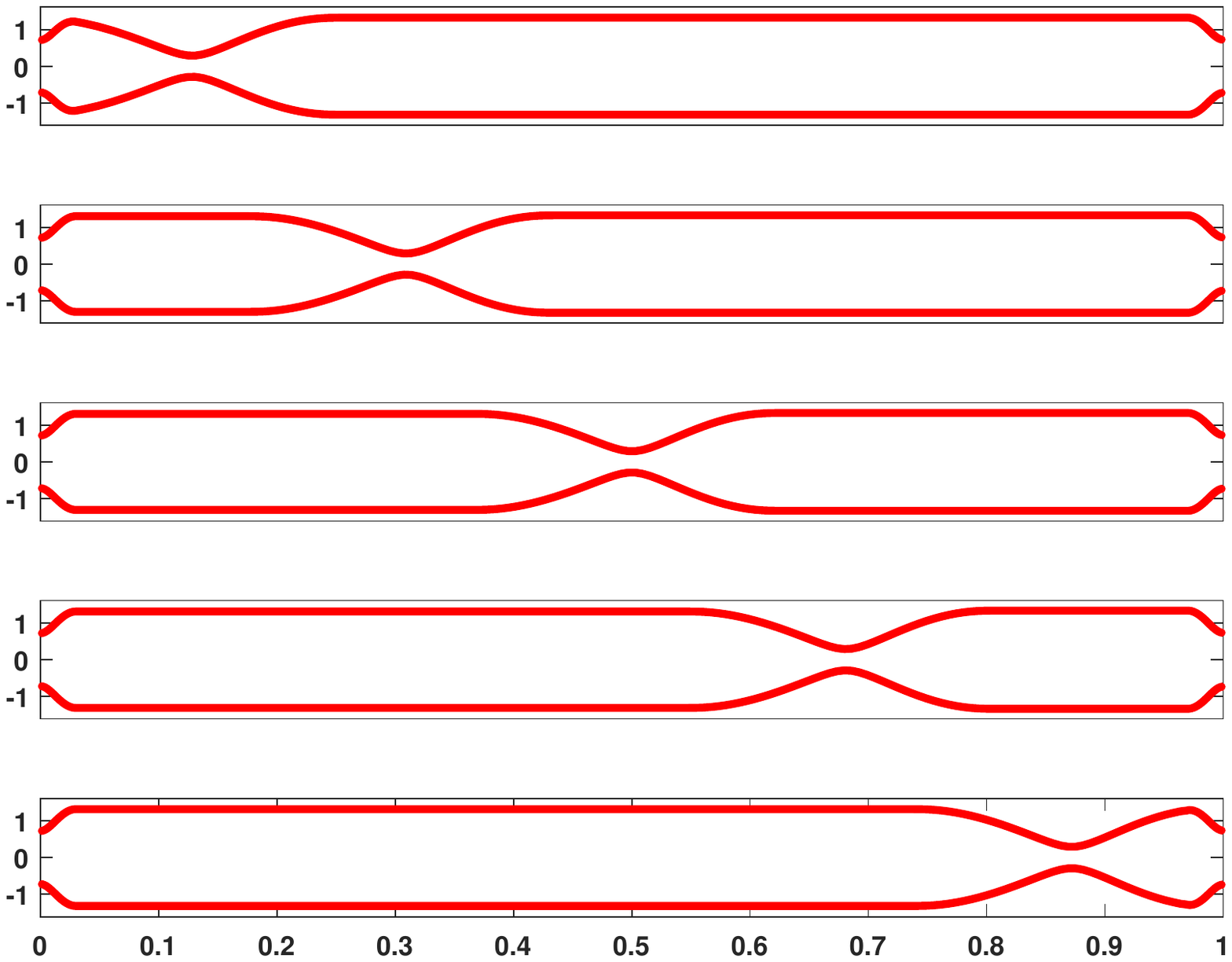}
        \caption{Regime 1}
        \label{fig:regime1_demo}
    \end{subfigure}
    \ 
    \begin{subfigure}[b]{0.24\textwidth}  
        \centering 
        \includegraphics[trim=75 210 60 210,clip,width=\textwidth]{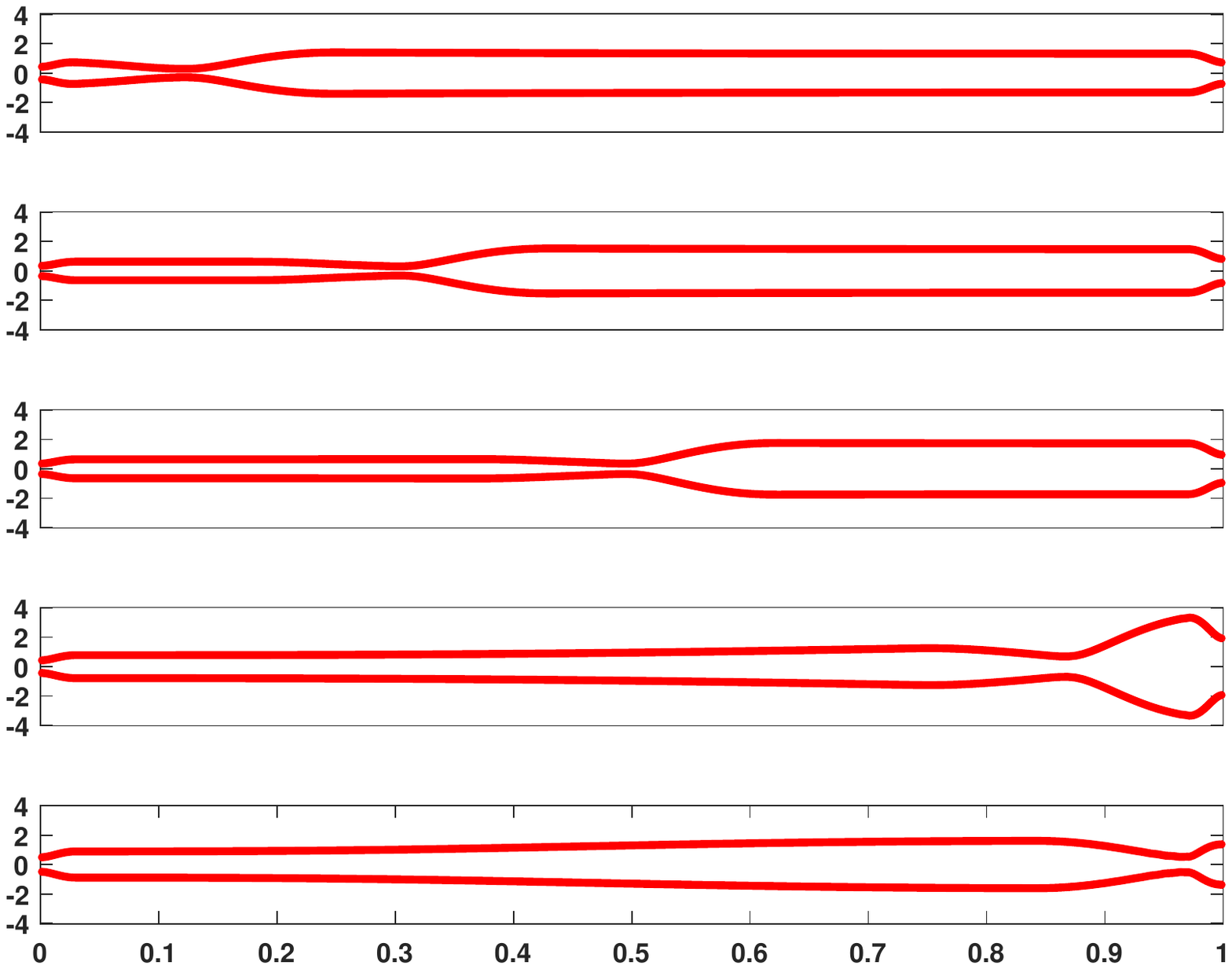}
        \caption{Regime 2}
        \label{fig:regime2_demo}
    \end{subfigure}
    \ 
    \begin{subfigure}[b]{0.24\textwidth}   
        \centering 
        \includegraphics[trim=75 210 60 210,clip,width=\textwidth]{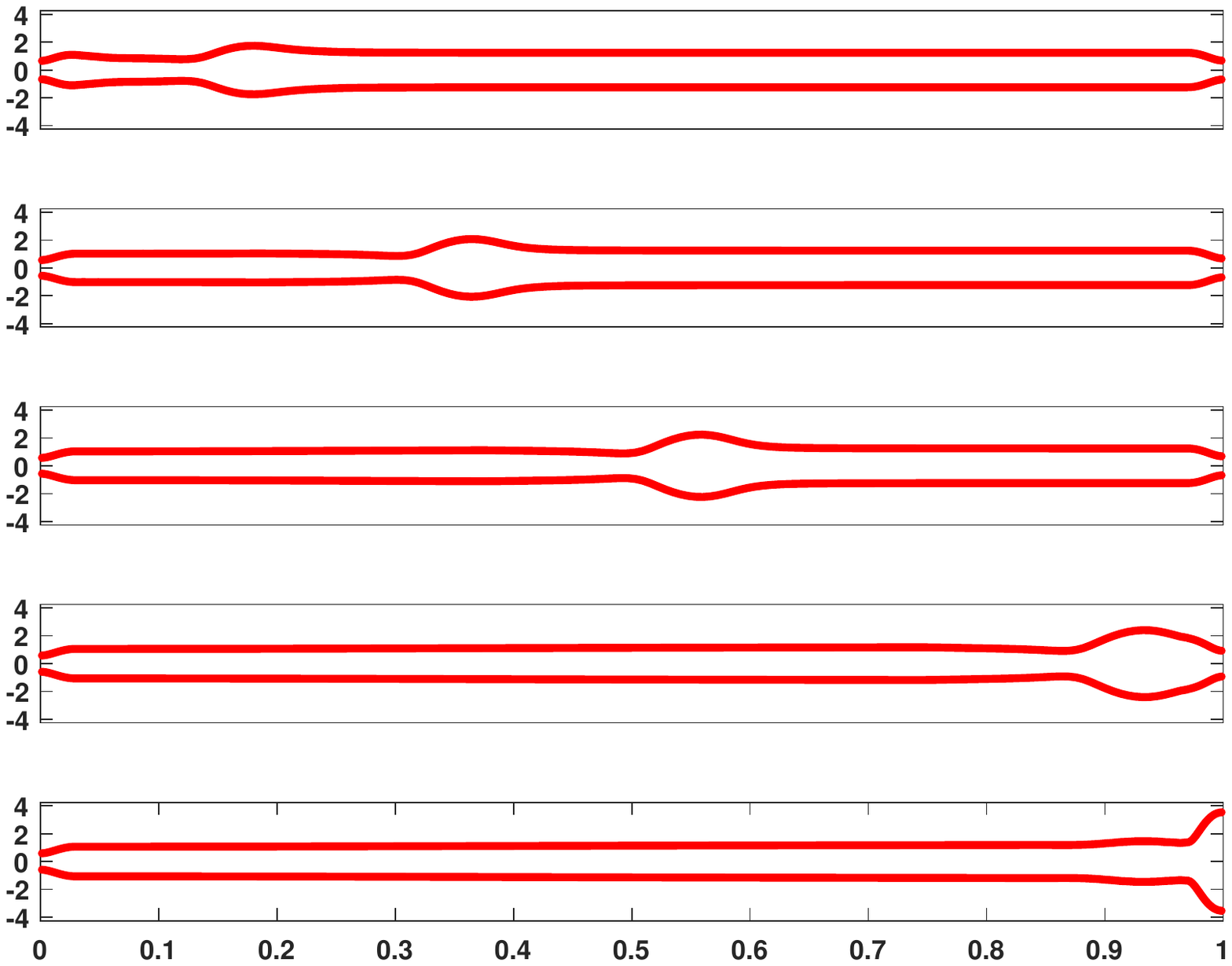}
        \caption{Regime 3}
        \label{fig:regime3_demo}
    \end{subfigure}
    \ 
    \begin{subfigure}[b]{0.24\textwidth}   
        \centering 
        \includegraphics[trim=75 210 60 210,clip,width=\textwidth]{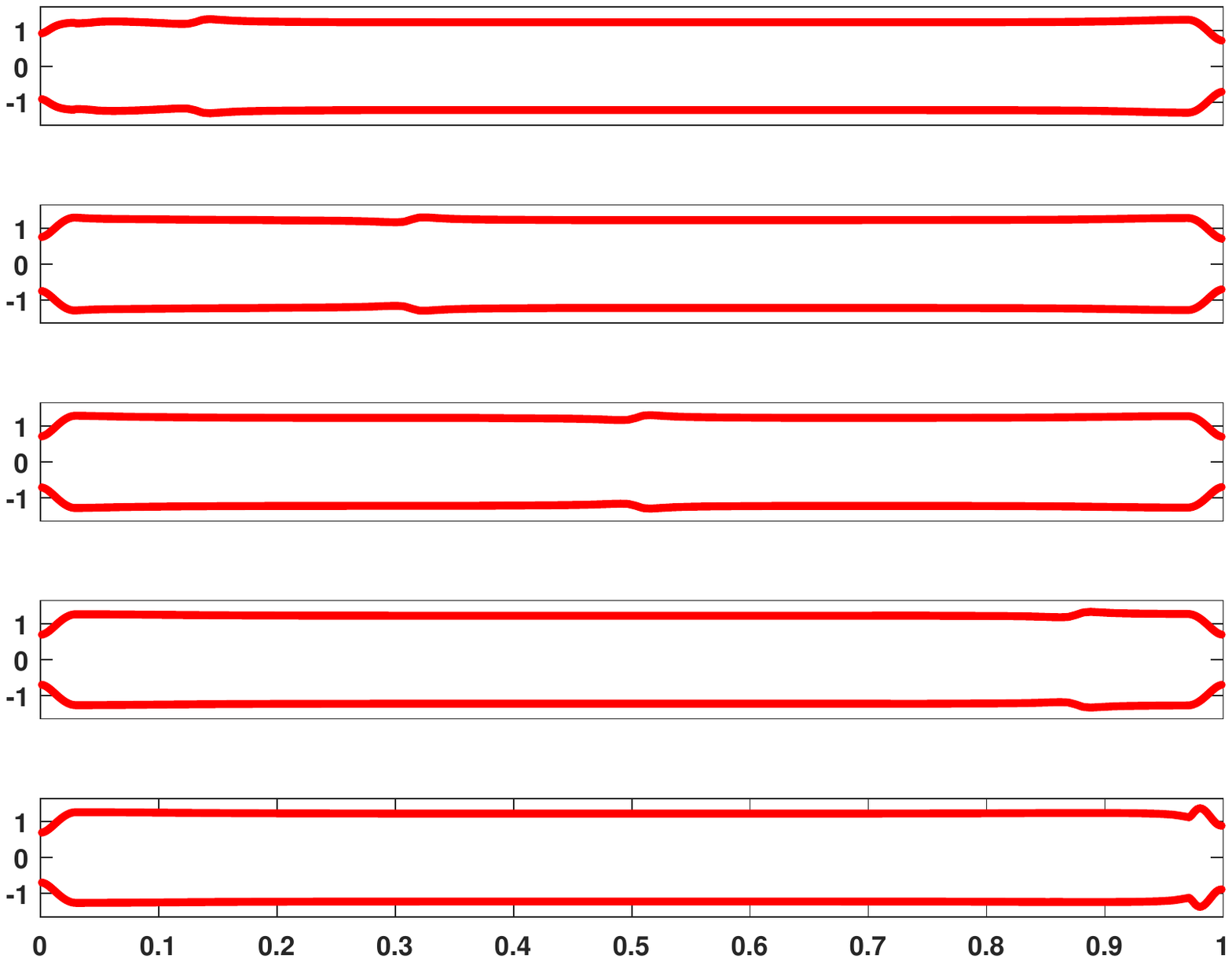}
        \caption{Regime 4}
        \label{fig:regime4_demo}
    \end{subfigure}
    \caption{Tube deformations corresponding to regimes occurring for physiologically
    relevant parameter ranges. Definitions used to classify a solution as
    belonging to a particular regime are given in Table~\ref{tab:regime_def}.}
    \label{fig:all_regimes_demo}
\end{figure*}

With all the parts of the 1-D model in place, we can now investigate the
system's response to the applied activation as a function of the operating
parameters, $\theta_0$, $\psi$ and $\beta$. Changing these values is
equivalent to investigating the effect of tube wall stiffness, wave speed,
contraction strength, fluid density and flow resistance on the tube wall
deformation and internal flow patterns during peristalsis. After studying the
system's behavior for a broad range of operating values, we found four
distinct, physiologically relevant patterns of peristaltic pumping based on
the way the tube walls and the fluid inside respond to the applied activation.
These regimes are shown in Fig.~\ref{fig:all_regimes_demo}. The progression 
of these regimes from numbers 1 to 4 can be interpreted simply as the response
of the system as fluid viscosity is continually increasing. It should be noted
that the fourth regime displays very little deformation of tube area.  The
deformation dynamics for this regime are unremarkable and the tube shape
remains quite similar to the shape of the initial condition for the entire
duration.

The occurrence of these regimes is dictated by a competition between elastic
forces generated due to deformation of the tube wall and resistance to flow
through the narrowest part of the contraction. A relatively stiff tube wall
resists deformation and forces the fluid that was displaced due to the
advancing wave to flow back through the contraction. In this scenario, we see
the walls deform as shown in Regime~1. An equivalent way of looking at this is
to assume a low resistance to flow through the contraction. The energy
required to expand the tube walls is significantly greater than the energy
required to overcome viscous resistance  across the contraction and thus, the
pumping is almost quasi-steady with the tube walls having the same diameter on
either side of the contraction. On the other hand, if the resistance to flow
is high, it is favorable for the system to expand the tube walls to
accommodate the fluid that is being displaced by the advancing peristaltic
wave. At moderate flow resistances, this exact situation occurs when the
stiffness of the tube walls is low. The tube deformation pattern in this
scenario is denoted as Regime~2.

Regime~3 has been investigated in great detail by Takagi and Balmforth \cite{Takagi2011}. In their
work, the lubrication approximation is used in an infinitely long domain with
open ends. In our configuration, we see the formation of a ``blister'' as
predicted by their model when the resistance to flow is high and the tube
walls are fairly compliant. The formation of a blister is due to the fact that
the amount of time it takes for the fluid to redistribute is comparable to the
time it takes for the wave to travel along the tube length. Regime~4 occurs
when the resistance to flow is much higher than that in Regime~3. In this
case, the amount of time it takes for the fluid to `move' and respond to the
peristaltic contraction is much higher than the amount of time it takes for
the wave to travel over a section of the tube. The fluid velocity in this case
is extremely low leading to a small value in $\partial A/\partial t$ as well.
In this region of operation, the fluid essentially behaves as a semi-solid
entity and the contraction causes only a small local deformation as it travels
forward.

{When the value of $\psi$ (which is the inverse of the Mach
number squared) is small, the wave-like nature of the system begins to emerge.
In this region the speed of the peristaltic wave is comparable to the speed at
which a disturbance in the system travels at. For the esophageal wall, this
disturbance speed is of the order of 1 m/s, but the peristaltic wave speed
rarely exceeds 10 cm/s. As such, we will not investigate the dynamics of the
system at these unphysiologic wave speeds.}


\section{System behavior represented as a regime map} \label{regime_map_analysis}

In Section~\ref{dynamics_1d}, four patterns of observed tube wall deformation
depending on the system's operating conditions were presented
(Fig.~\ref{fig:all_regimes_demo}). Our goal was to combine the various
operating parameters to summarize the response of the system in the form of a
regime map. We wished to find the least number of parameters that could be
used to describe the response of the system to the peristaltic contraction
wave. Using the Buckingham-Pi approach leads to the several non-dimensional
numbers already shown above. Plotting the system's response for each of these
quantities is cumbersome. To that effect, we utilized the wave-frame approach
that has often been employed by other \textcolor{REDCOLOR2}{researchers} to analyze peristaltic flow
to find the most convenient combination of operating parameters to visualize
the system's response. \textcolor{REDCOLOR2}{Detailed derivation of these parameters is provided in
the Supplemental Materials section on the ASME Digital Collection.}

\begin{figure}
    \centering
    \fbox{\includegraphics[trim=160 150 0 105,clip,width=0.5\linewidth]{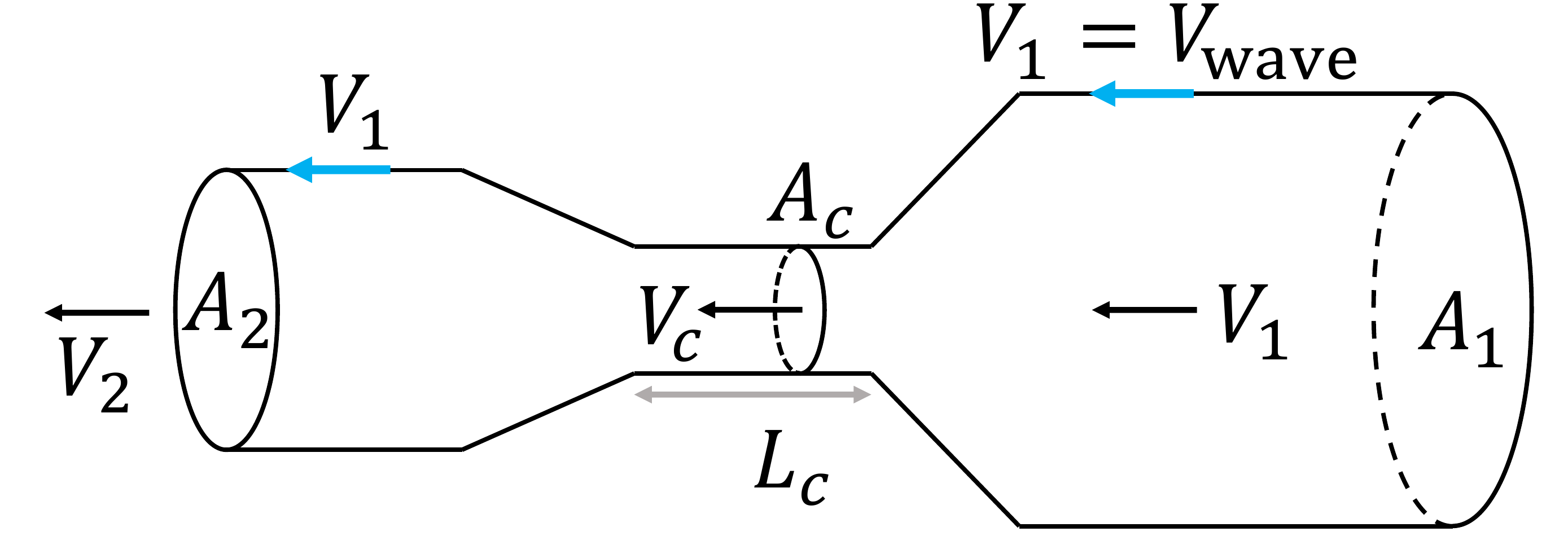}}
    \caption{Peristaltic contraction as observed in the wave's frame of reference}
    \label{fig:wave_frame_cz}
\end{figure}

\textcolor{REDCOLOR2}{From the wave frame-based analysis based on Fig.~\ref{fig:wave_frame_cz}, we systematically combined the tube
stiffness coefficient, contraction properties, fluid viscosity and peristaltic
wave speed to obtain two nondimensional numbers: cumulative flow resistance $\psi_{f}^{\prime}$ (or
$\Psi_{f}^{\prime}$) and cumulative stiffness $\psi_{k}^{\prime}$ that were used to describe the
system's response to a peristaltic contraction.  Depending on the specific version of the viscous term
used (parabolic flow or friction factor-based) in Eq.~(\ref{eq:momentum}), $\psi_{f}^{\prime}$ or $\Psi_{f}^{\prime}$ is
computed from the input parameters.} The two terms $\psi_{k}^{\prime}$ and $(\psi_{f}^{\prime},\
\Psi_{f}^{\prime})$ together, account for all the relevant parameters of the
system. The  effect of tube stiffness, fluid density, wave velocity and the
amount of fill volume of the tube is accounted for in $\psi_{k}^{\prime}$ and
$\psi_{f}^{\prime}$ accounts for the fluid's resistance to flow in the tube,
the length of the contraction zone and its intensity. All of these effects
contribute to the increase or decrease of the area ratio $A_2/A_1$ and hence the observed regime. {The
utility of our analysis is now clear because if we wished to visualize the
effect of these variables on pumping patterns, we would either need several
figures or utilize a plot that has multiple axes, corresponding to wave
velocity, tube stiffness, fluid viscosity, etc. However, by combining these
parameters in a sensible manner, made possible by the wave-frame analysis, the
visualization of observed regimes can be achieved in a clear and coherent
way.} The final combination of operating parameters used to compute these
cumulative nondimensional numbers is

\begin{equation}  \label{eq:pkp_eqn}
\psi_k^\prime = 2\psi\alpha_\mathrm{IC} = 2\left(\frac{K}{\rho c^2}\right)\alpha_\mathrm{IC}
\end{equation}

\begin{equation}  \label{eq:pfp_eqn}
\psi_f^\prime = 2\beta w \left(\frac{1-\theta}{\theta^2}\right) \quad \mathrm{for\ Low\ Re\ with\ \ } \beta = \frac{8\pi\mu L}{\rho A_0 c}
\end{equation}

\begin{equation}  \label{eq:PFP_eqn}
\Psi_f^\prime = 2\beta w \left(\frac{1}{\sqrt \theta}\right)\left(\frac{1-\theta}{\theta}\right)^2  \quad \mathrm{for\ High\ Re\ with\ \ } \beta = \frac{f L}{2 D_0}
\end{equation}

\begin{figure}
\centering
\includegraphics[trim=80 180 120 217,clip,width=0.5\linewidth]{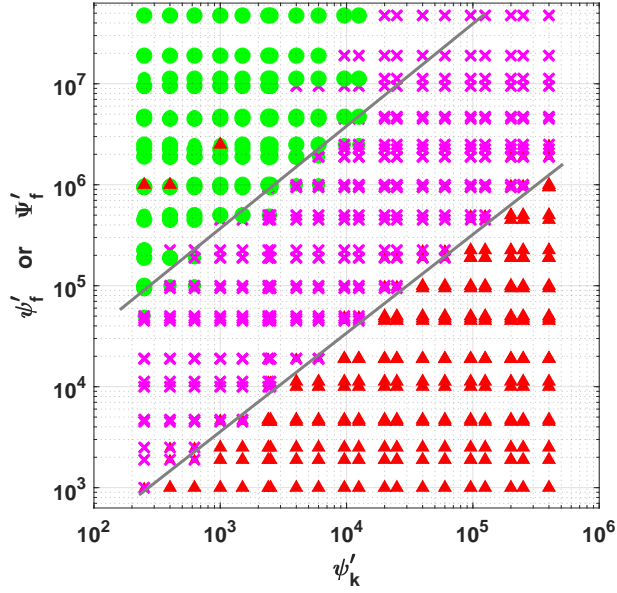}
\caption{Regime map showing system's response for a variety of input conditions for 
both parabolic and friction factor-based flow resistance terms. The parameters
$\psi_f^\prime$  and $\psi_k^\prime$ are computed using Eqs. (\ref{eq:pkp_eqn}),
(\ref{eq:pfp_eqn}) and (\ref{eq:PFP_eqn}) for  each data point. The x-axis
represents a cumulative stiffness parameter and the y-axis represents a
cumulative flow resistance parameter. Legend: Regime 1 ($\color{red} \blacktriangle$),
Regime 2 ($\color{my_purple} \pmb{\times}$), Regimes 3,4
({\Huge$\color{green} _\bullet$})}
\label{fig:Regime_map}
\end{figure}

{The system's response was investigated for a wide range of physiologically
relevant parameters. Values of $\psi$ were selected from the following set:
\{0.1, 0.25, 0.6, 1, 2.4, 5, 10, 24, 50, 100\}$\times 10^3$ and $\beta$ had
the following values: \{0.1, 0.25, 1, 5, 10, 50, 100, 250\}$\times 10^3$. Two
values of the initial condition $\alpha_{IC}$=\{1.25, 2.0\}, corresponding to
the volume of fluid in the tube, were chosen. The following values for
contraction intensity $\theta_0$ were selected: \{0.05, 0.1, 0.15, 0.2\}. {The
nondimensional width $w$ was set to 0.25 for all runs.} All possible
combinations of these parameters leads to 640 cases each for the parabolic and
friction factor-version of the momentum equation.} {The damping parameters and
solver tolerance were set to the same values that were used during
verification using the Method of Manufactured Solutions. Volume of fluid in
the tube was monitored as a measure of solution accuracy. Percent change in
tube volume across all runs was of the order of $10^{-11}$.} {The tube shape
computed from each of these runs was analyzed when the contraction was at the
halfway point i.e. at $\chi=0.5$. At this instant, the ratio $A_2/A_1$ was
computed along with the maximum tube area achieved ($A_\mathrm{max}$) and area
of the contraction ($A_c$). Based on visual inspection of their values for
several cases, the tube shape was then classified as Regime 1, 2 or 3/4 using
the definitions provided in} Table~\ref{tab:regime_def}. {Complete parametric
information and the associated regime for each case plotted in the regime map
is provided in a data sheet available in the Supplemental Materials on the
ASME Digital Collection.}

\begin{table}[t]
\caption{Definitions used to classify each solution as belonging to a particular regime \label{tab:regime_def}}
\centering{
\renewcommand{\arraystretch}{1.25}
\begin{tabular}{l l l}
\toprule
Regime 1 & & ${A_1}/{A_2}<1.7$               \\
Regime 2 & & ${A_1}/{A_2}\geq1.7$            \\
Regime 3 & & $A_{\mathrm{max}}/{A_1} > 1.5$  \\
Regime 4 & & $A_c/A_2 > 0.9$                 \\
\bottomrule
\end{tabular}
}\\

{\justify Note: Areas $A_1$ and $A_2$ are the distal and proximal areas (Fig.~\ref{fig:wave_frame_cz})
 and $A_\mathrm{max}$ is the area of
the bulge (observed in Regime 3 only) at its widest point. For Regimes 1 and
2, $A_\mathrm{max}=A_1$. The area at the contraction is denoted by $A_c$. \par}
\end{table}

Armed with this simplification, we present the occurrence of each regime as a
set of points on a $\psi_{f}^{\prime}$ vs. $\psi_{k}^{\prime}$ plot.
Fig.~\ref{fig:Regime_map} shows the occurrence of the various regimes at
different values of the two $\psi^\prime$s. We combine all the regime points
from the two different flow types and present them in a single plot. We see
that the regime data points from both flow types fall within the general
vicinity of each other. The set of points belonging to each regime is
represented by a specific marker. 

{The wave frame approach that was taken to find the combination of parameters
does not account for the formation of a bulge or a blister. As both regimes 3
and 4 occur in the region of high fluid viscosity, they are combined into a
single entity for plotting purposes.}  From this plot, we clearly observe the
presence of a general boundary between each set of regime points. {The
boundary represents the change of the system from one solution to another.
This analysis can be used as a starting point to quantitatively derive the
location of the boundaries between the regimes shown in
Fig.~\ref{fig:Regime_map}. In addition to the parameters mentioned above, it
has been observed that there is active relaxation ahead of, and stiffening
behind, the contraction wave that can change the effective material properties
of the tube to improve bolus transport through the esophagus
\cite{Mittal2016, AbrahaoJr2010}. In a future work, we will be incorporating
these non-uniform changes in effective material properties into our analysis
and study the effect on the observed regimes and quantitatively predict the
regime boundaries as a function of all physiologically relevant parameters.}

{With the help of the regime map, one can predict the shape
assumed by the tube during peristalsis by simply computing the relevant
$\psi_{f}^{\prime}$ and $\psi_{k}^{\prime}$ from the given operating
conditions and  finding the region which these points belong to. The regime
map also helps in understanding the change in tube shape that would occur as
one of the operating conditions is changed. For instance, as the tube
stiffness increases, the system will deform in such a way that the area ratio
 tends to 1.0. Increasing the fill volume or decreasing the
peristaltic wave velocity leads to a similar outcome. When it comes to
increasing the fluid's viscosity, we see a transition from Regime 1 to Regime
2 as predicted by the regime map when the value of $\psi_f^{\prime}$ is
increased. When the area contraction factor $\theta$ is reduced i.e. the
amount of wave `squeezing' is increased, we again observe the system
transition from Regime 1 to Regime 2.}

{Each of these regimes correspond to a specific deformation
pattern observed using the FLIP device in different subjects and/or diseases.
For instance, assume that the device is calibrated and the operating
conditions are benchmarked in such a way that it shows pumping  patterns
corresponding to Regime 2 in a healthy individual. Under the same operating
conditions, a patient with  peristaltic dysfunction will display pumping
patterns corresponding to either Regime 1 or Regime 3/4. Depending on the
specific behavior displayed, the regime map helps us identify the cause of the
abnormality. For instance, a patient displaying regime 1 might have a stiffer
esophagus due to fibrosis and a patient with Regime 4 has dysphagia due to
ineffective peristalsis. Thus the quantification of the device's behavior in
the form of these regime maps directly assists in better interpreting the
various shapes seen in patients during the FLIP procedure.}


\section{Defining and quantifying pumping effort}

In this section, we aimed to understand how energy is spent during a
peristaltic contraction in a closed tube. Finding the work done by a
peristaltic wave and observing its variation during pumping can offer further
insights into the pumping process. The configuration that is being analyzed in
this work is a closed system. As such, there is no ``net flow rate'' or
non-zero displacement of fluid volume during one peristaltic event which can
be used to quantify efficacy. At the end of peristalsis, all the work done by 
the peristaltic wave is dissipated due to fluid viscosity. Hence, the total
pumping work done is simply the amount of energy that has dissipated. During
peristalsis however, the stretching of the esophageal walls leads to some of
the energy being stored as elastic potential energy. After passage of the
peristaltic wave, the tube relaxes and releases the energy back into the fluid
which is then lost via viscous dissipation. To understand the quantitative
relationship between these three agents, we turn to the parabolic version of
the  momentum equation (\ref{eq:momentum}) and multiply both sides with $Au$
to obtain

\begin{equation} \label{eq:momAu}
  \rho Au\frac{\partial u}{\partial t}+\rho\left(Au\right)u\frac{\partial u}{\partial x} = 
  -Au\frac{\partial p}{\partial x}-\frac{8\pi\mu Qu}{A}.
\end{equation}

Noting that $Q=Au$, we rewrite the equation to form terms that involve
derivatives of the kinetic energy of the fluid per unit volume. We end up with

\begin{equation}
A\frac{\partial}{\partial t}\left(\frac{1}{2}\rho u^{2}\right)+ 
Q\frac{\partial}{\partial x}\left(\frac{1}{2}\rho u^{2}\right)=
-Au\frac{\partial p}{\partial x}-\frac{8\pi\mu uQ}{A},
\end{equation}

\noindent to which we add the continuity equation (multiplied with $\rho u^{2}/2$) on
the LHS  and after combining terms using the product rule, we end up with the
equation

\begin{equation} \label{eq:midstep}
\frac{\partial}{\partial t}\left(\frac{1}{2}\rho Au^{2}\right)+ 
\frac{\partial}{\partial x}\left(\frac{1}{2}\rho Qu^{2}\right)=
-Au\frac{\partial p}{\partial x}-\frac{8\pi\mu uQ}{A},
\end{equation}

\noindent the terms of which can be interpreted as follows,

\begin{equation}
\frac{\partial}{\partial t}\left(\frac{1}{2}\rho Au^{2}\right) +
\frac{\partial}{\partial x}\left(\frac{1}{2}\rho Qu^{2}\right) = 
-\underbrace{\frac{\partial\left(Aup\right)}{\partial x}}_{\text{surface work}}
+\underbrace{p\frac{\partial\left(Au\right)}{\partial x}}_{\sim\ p\nabla\cdot\mathbf{u}}
-\underbrace{\frac{8\pi\mu uQ}{A}.}_{\substack{\text{viscous}\\\text{dissipation}}}
\end{equation}

\noindent Note that conservation of volume can be used to replace the
$\partial\left(Au\right)/\partial x$ term with $-\partial A/\partial t$ to
obtain

\begin{equation} \label{eq:start_point}
\frac{\partial}{\partial t}\left(\frac{1}{2}\rho Au^{2}\right) + 
\frac{\partial}{\partial x}\left(\frac{1}{2}\rho Qu^{2}\right) = 
-\frac{\partial\left(Aup\right)}{\partial x}-p\frac{\partial A}{\partial t}-\frac{8\pi\mu uQ}{A}.
\end{equation}

Now we integrate this equation over the length of the tube from $x=0$ to $L$
and apply the zero velocity condition $Q=u=0$, at the tube ends. This leads to
terms with $\partial x$ going to zero giving

\begin{equation}  \label{eq:energy_balance_no_split}
-\int\limits _{0}^{L}p\frac{\partial A}{\partial t}\mathrm{d}x =
\frac{\partial}{\partial t}\int\limits _{0}^{L}\left(\frac{1}{2}\rho Au^{2}\right)\mathrm{d}x +
\int\limits _{0}^{L}8\pi\mu u^{2}\mathrm{d}x.
\end{equation}

\noindent Equation~(\ref{eq:energy_balance_no_split}) succinctly shows how power is
distributed in the system at each time instant. Without the negative sign, the
pressure term represents the work done by the fluid on the tube walls. Thus,
the LHS is the work done by the walls on the fluid. The first term on the RHS
is the rate of change in the kinetic energy of the fluid and the last term is
the rate of energy loss due to viscous dissipation. It is important to realize
that the pressure term has contributions from both the passive elastic part of
the esophageal walls and the rise due to active contraction. When separated,
we get a better understanding of the power breakdown, 

\begin{equation}
\label{eq:power_breakdown_passactv} -\int\limits
_{0}^{L}p_{\text{actv}}\frac{\partial A}{\partial t}\mathrm{d}x = 
\frac{\partial}{\partial t}\int\limits _{0}^{L}\left(\frac{1}{2}\rho
Au^{2}\right)\mathrm{d}x \ + \int\limits
_{0}^{L}p_{\text{pass}}\frac{\partial A}{\partial t}\mathrm{d}x + \int\limits
_{0}^{L}8\pi\mu u^{2}\mathrm{d}x. 
\end{equation}

Simply put, the LHS is the rate of work done by the active part of the tube
wall (the peristaltic contraction) on the confined  fluid and the terms on the
RHS show the consumers of this spent power. Some of it goes into increasing
the kinetic energy of the fluid, some of it is stored in the tube walls when
they stretch due to an increase in local pressure and the rest is lost due to
viscous dissipation. At this point, we need to assume some
form of the passive pressure to compute the values of each these terms for the
various regimes displayed in our 1D model. The most logical form of the
passive pressure component is a linear dependence on the tube area as seen in
\cite{Kwiatek2011}. The breakdown of the fluid pressure can then be trivially
split as
\begin{equation}
P=\frac{\alpha}{\theta}-1 = 
\underbrace{\left(\alpha-1\right)}_{\text{passive}} + 
\underbrace{\alpha\left(\frac{1}{\theta}-1\right)}_{\text{active}}.
\end{equation}
Using this breakdown of fluid pressure, we can compute the values of each of
the terms in Eq. (\ref{eq:power_breakdown_passactv}) and visualize the
variation of work done or energy dissipated over a single peristaltic event
for each regime. The above analysis can be repeated for the momentum equation
with the friction factor stress term but the resulting work variations show no
qualitative differences between the two scenarios.

\begin{figure*}
\begin{subfigure}[c]{0.32\textwidth}
\centering{\includegraphics[trim=90 240 110 240,clip,width=\textwidth]{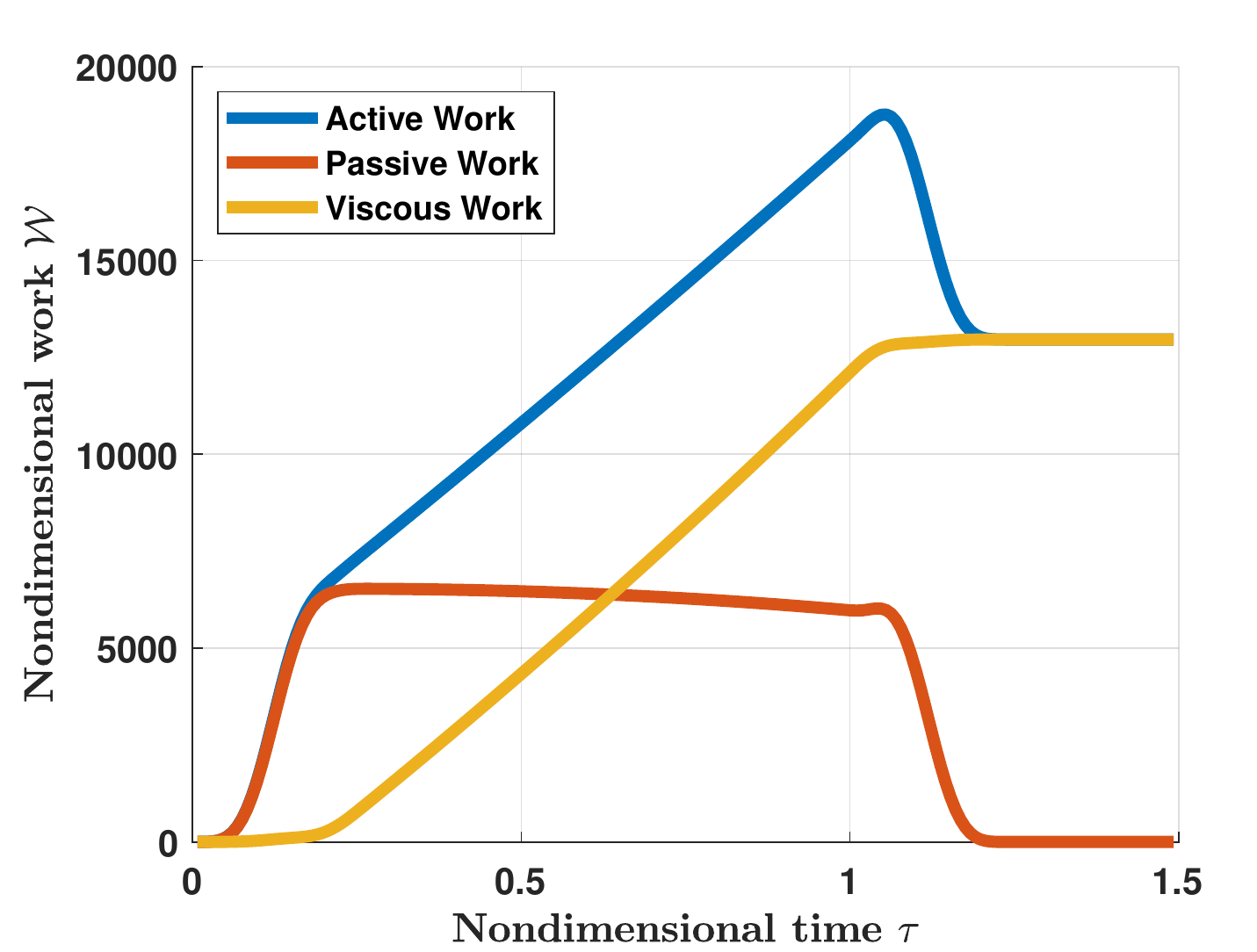}}
\subcaption{Regime 1}
\end{subfigure}
\begin{subfigure}[c]{0.32\textwidth}
\centering{\includegraphics[trim=90 240 110 240,clip,width=\textwidth]{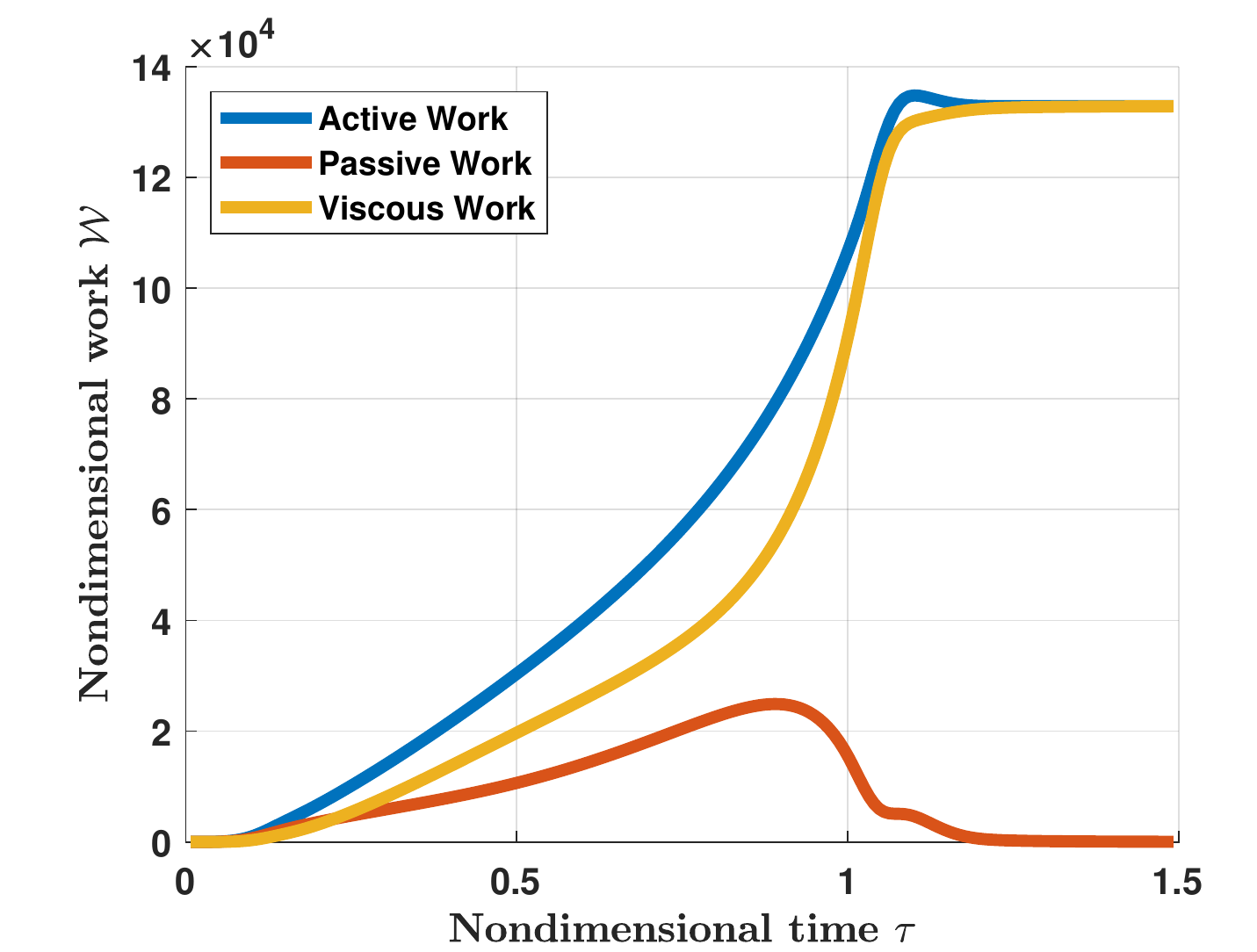}}
\subcaption{Regime 2}
\end{subfigure}
\begin{subfigure}[c]{0.32\textwidth}
\centering{\includegraphics[trim=90 240 110 240,clip,width=\textwidth]{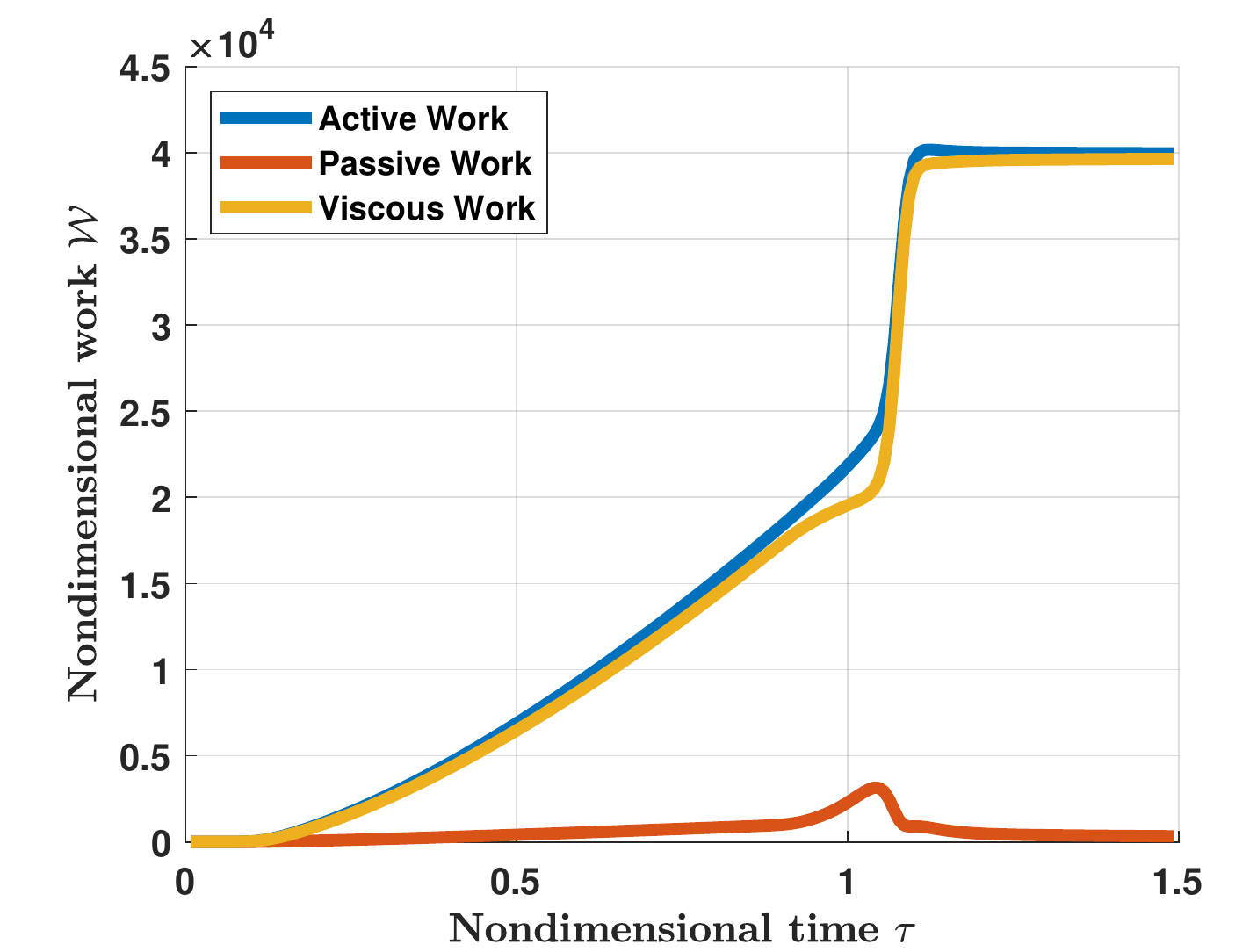}}
\subcaption{Regime 3 \label{fig:regime3_work_curves}}
\end{subfigure}
\caption{Representative work curves for regimes that show non-trivial tube
deformation. Each regime shows a unique signature for variation of work during peristalsis. At each instant, the sum of passive and viscous work is
equal to the active work done.}
\label{fig:all_work_curves_laminar}
\end{figure*}


\subsection{Work curves from the reduced order model} \label{workcurves1dtheory}

As explained earlier, Eq. (\ref{eq:power_breakdown_passactv}) gives the
balance of power at each instant of time. Integrating this equation  over time
gives the balance of work done or energy lost as the peristaltic wave
advances. In Fig. (\ref{fig:all_work_curves_laminar}), we show the energy 
contribution from each of these terms {{for regimes with
non-trivial tube shapes}}. This figure summarizes how the active work is split
between the two sinks over the entire range of observed pumping patterns. At
time $\tau=0$, the peristaltic wave begins traveling over the tube length and
the active work done by it is zero. As the wave advances, the work done by it
i.e. the active work is split into either increasing the potential energy
stored in the tube walls or generating flow fields that then lose energy via
dissipation. {For the parameter space that is being analyzed in this work
($\psi\gg1$), the rate of change of fluid kinetic energy is negligible and has
not been plotted. Non-dimensionalizing}
Eq.~(\ref{eq:power_breakdown_passactv}) {results in}

\begin{equation} \label{eq:power_breakdown_passactv_nondim}
-\psi\int\limits
_{0}^{1}P_{\text{actv}}\frac{\partial\alpha}{\partial\tau}\mathrm{d}\chi = 
\frac{\partial}{\partial\tau}\int\limits _{0}^{1}\left(\frac{1}{2}\alpha
U^{2}\right)\mathrm{d}\chi +   \psi\int\limits
_{0}^{1}P_{\text{pass}}\frac{\partial\alpha}{\partial\tau}\mathrm{d}\chi +
\beta\int\limits _{0}^{1}U^{2}\mathrm{d}\chi,
\end{equation}

\noindent {which shows that for large values of $\beta$ and $\psi$ (as is the case for
flows relevant to esophageal peristalsis), the term corresponding to rate of
change of fluid kinetic energy is very small compared to the other terms in
the energy balance equation.}

The work curves for each of these regimes show unique identifying features
that confirm what is visually observed through the tube deformation patterns.
For Regime 1, once the wave has created a zone of reduced area, the active
work goes into overcoming the viscous resistance and the potential energy stored in
the walls remains unchanged. Upon approaching the right boundary, the
contraction leaves the domain causing the tube to relax and the stored elastic
energy is recovered. For Regime 2 however, we observe a gradual rise in stored
elastic energy with time indicating that the activation wave is continuously
doing work on the tube wall. Unlike Regime 1, when the wave approaches the end
and allows the tube walls to relax, the stored energy is lost via fluid
dissipation. This is observed by following the viscous work curve which
sharply rises around the $\tau=0.8$ mark to meet the active work curve. It is
also interesting to note that in spite of large  tube wall deformations
associated with this regime, the majority of active work done is still lost
via viscous dissipation. Work curves for Regime 3 reflect the low wall
deformations observed from the tube shape plots which only show a bulge ahead
of the contraction. The passive work done is small compared to the  active
work done which is almost entirely lost via viscous dissipation. {When the
contraction approaches the end, it acts on the bulge in the tube ahead of it
as it cannot go any further. Contraction of this bulge then reduces its area
and causes the accumulated fluid to flow in the opposite direction. This leads
to the sharp rise in viscous work at $\tau=1$ as seen in}
Fig.~\ref{fig:regime3_work_curves}. For Regime 4 there is a simple, linear
increase in viscous dissipation with time. Passive work done is zero for all
time due to negligible wall deformation.

This detailed look into the variation of the active, passive and viscous work
done during a peristaltic event gives us the tools to identify what a healthy
pumping wave looks like and the conditions under which one might observe
reasonable tube wall deformations. A significant change in the tube area for
this configuration due to peristalsis (as seen with the healthy peristaltic
wave in Fig.~\ref{fig:FLIP_5MT_example}) indicates that the wave has some
ability to move fluid forward in a normal setting which involves bolus
transport following normal swallowing.

\subsection{Work curves from FLIP patient data}

In the previous subsection, we have defined and provided a strong mathematical
foundation for computing work done during peristalsis in the current
configuration and the various sinks that consume the energy generated by
muscular activation. The work calculations were performed in the context of
the reduced order model presented in Section \ref{oneDmodeldetails}. In this
section, we wish to extend the utility of the model to compute work curves
using data obtained from the FLIP device when used in human subjects. The goal
was to get a sense of the magnitude of peristaltic work done in normal
subjects and to see what characteristics are observed in work curves obtained
from patient data.

{ It is clear that the 1D model in Section
\ref{oneDmodeldetails} takes as input the activation wave and predicts the
resultant tube wall deformation and the associated fluid velocity and pressure
fields. The FLIP data on the other hand contains the variation of tube wall
deformation as a function of time and the value of pressure at the
(approximate) location $x=\chi L = 16\ \mathrm{cm}$. Simply speaking, $A(x,
t)$ and $p(L, t)$ are already known. The values of fluid velocity and pressure
at all the other locations along the tube are unknown. The device measures the
lumen profile and stores diameter readings from each of the 16 sensors (as
seen in Fig.~\ref{fig:EndoFLIP_catheter_closeup}). The recorded diameters are
used to compute lumen area after accounting for the catheter's presence and
are corrected to ensure volume conservation in the tube. Equation
(\ref{eq:continuity}) is then used to calculate $U(x,t)$ from planimetry data
$A(x,t)$ collected by the device. With the available values of $A(x,t)$ and
$U(x,t)$, the momentum equation (\ref{eq:momentum}) is used to compute the
pressure gradient. Using $p(L,t)$, the pressure is then found as a function of
$x$ and $t$ for the entire duration of peristalsis. Implementation details of
this step using Eqs. (\ref{eq:continuity_nondim}) and (\ref{eq:momentum_nondim})
are provided in Ref.~\cite{Halder2021}}.

\begin{figure*}
    \centering
    \begin{subfigure}[b]{0.475\textwidth}
        \centering
        \fbox{\includegraphics[trim=115 230 115 252,clip,width=\textwidth]{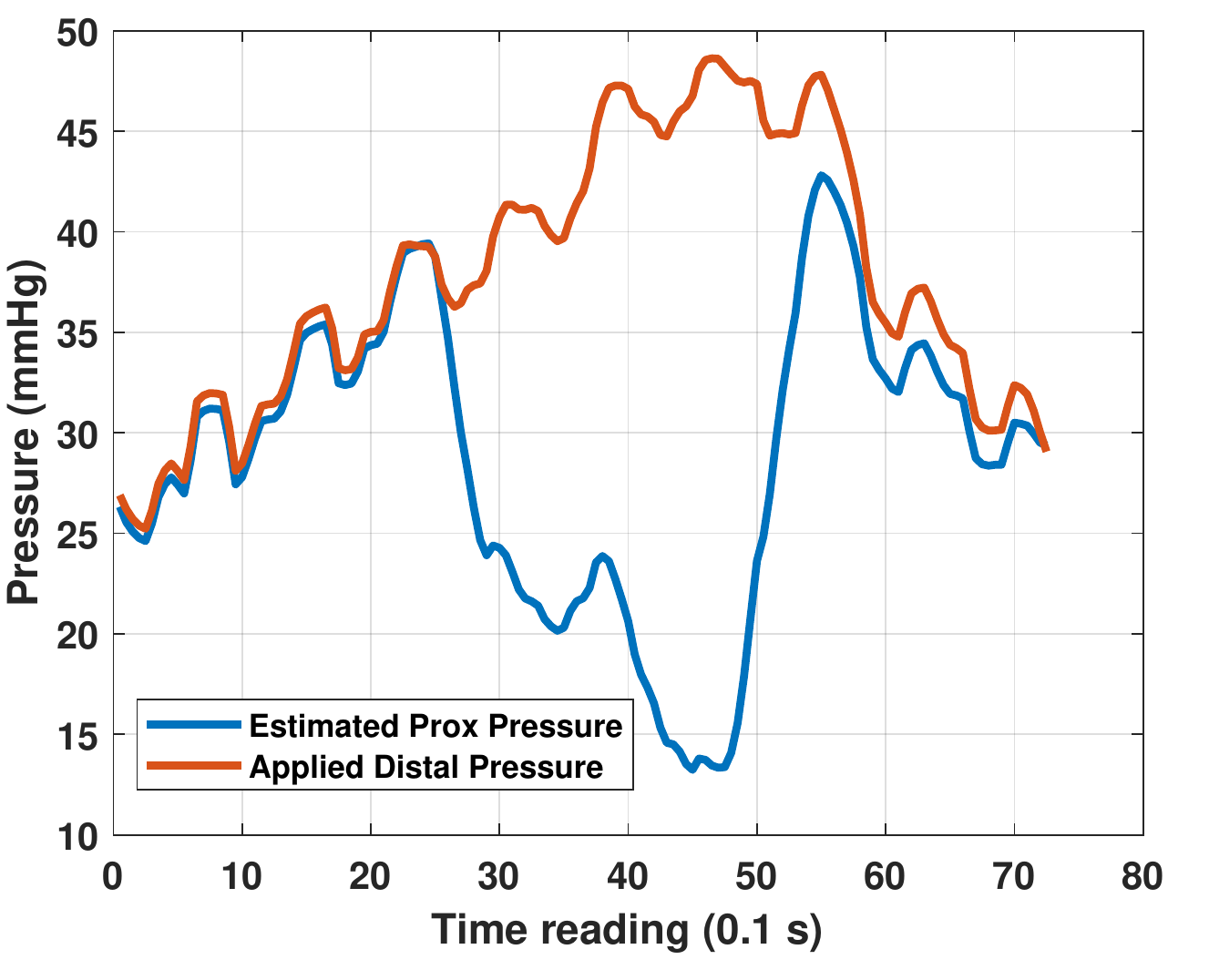}}
        \caption{Contraction 1 - pressure variations}
        \label{fig:sw1pres}
    \end{subfigure}
    \hfill
    \begin{subfigure}[b]{0.475\textwidth}  
        \centering 
        \fbox{\includegraphics[trim=100 230 130 252,clip,width=\textwidth]{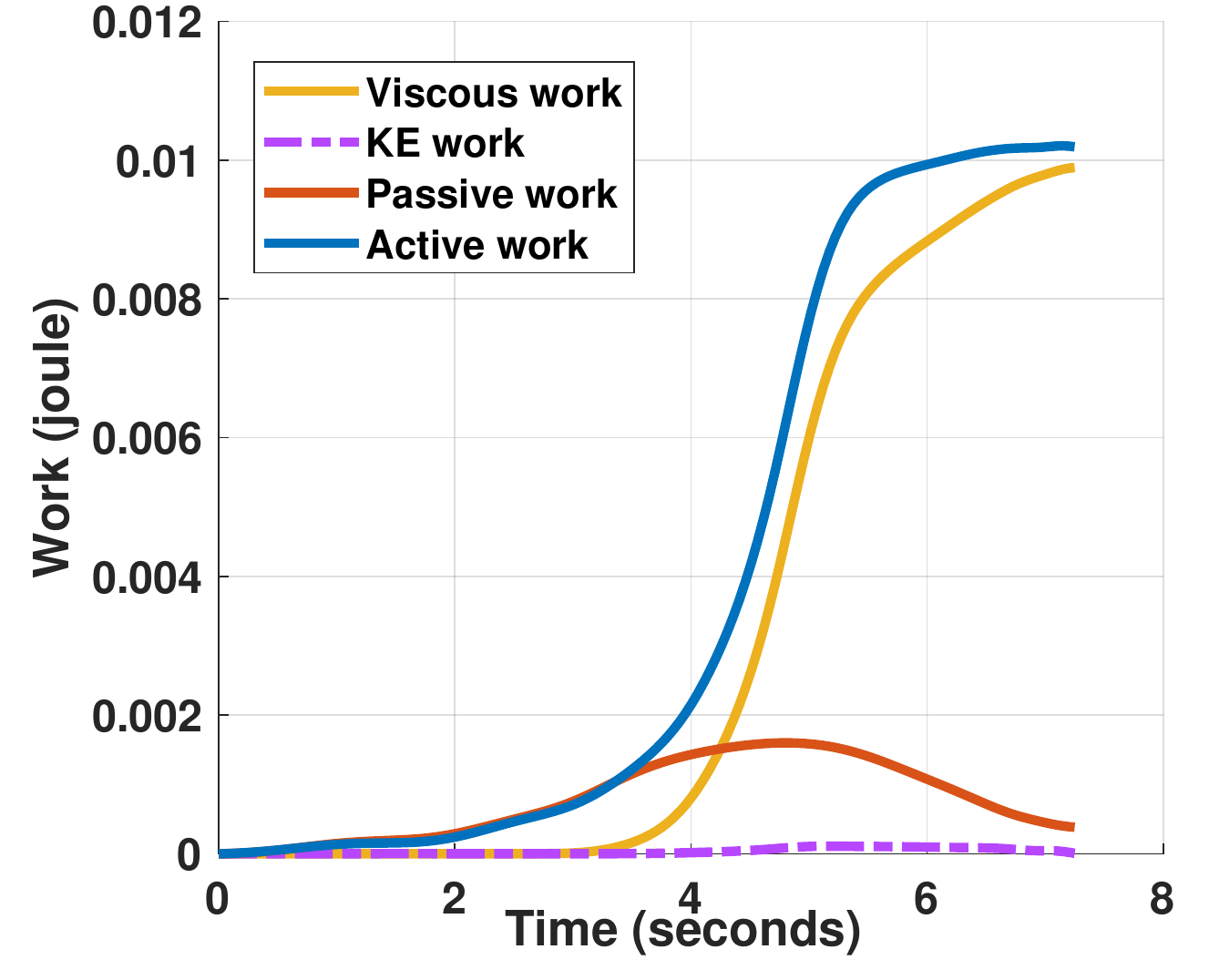}}
        \caption{Contraction 1 - work vs. time curves}
        \label{fig:sw1wrk}
    \end{subfigure}
    \caption{Estimated proximal pressure variation and work computations from FLIP data
    corresponding to a single peristaltic contraction occurring at 30 mL in a
    healthy control. The esophagus does not return to its exact starting shape
    at the conclusion of peristalsis leading to a small non-zero value of
    passive work at the end.} 
    \label{fig:two_swallows_work_pressure}
\end{figure*}

With these steps in place, we isolate a contraction wave to analyze and
generate work curves for. An example of this can be seen in Fig.
(\ref{fig:FLIP_5MT_example}). A window of readings is chosen that includes a
single pressure peak. At the start of computation, the wave begins to travel
over the tube length and results in a pressure rise. The contraction travels
along the esophagus as time goes on and by the last reading the wave has
finished traveling over the entire tube length and the computation ends. In
Fig. (\ref{fig:two_swallows_work_pressure}), we show the work curves and the
pressures at $\chi=0.2$ (referred to as the \textit{proximal} pressure) and
$\chi=1$ (referred to as the \textit{distal} pressure) as a function of time
for a typical peristaltic contraction. Again, we emphasize that the former is
predicted by our model and the latter is measured by the device and is applied
to the model to fix the pressure levels inside the tube.

The first thing to note from the analysis of the patient's secondary
peristaltic contractions is the pressure predicted by the model at the
proximal (left) end of the tube at $\chi=0.2$. When the wave is incoming, the
pressure in the entire system rises, but once it passes over the left end,
there is a sharp drop in pressure. This is due to the temporary displacement
of fluid due to the peristaltic pumping action of the wave leading to the tube
having reduced area at the left end. The displaced fluid then stretches the
walls of the distal end of the tube at $\chi=1$ and this is marked by the
continuous rise in pressure at that location. Once the wave has passed, the
fluid accumulated at the end flows back and the tube attains a uniform shape
and pressures at both locations equalize at the end of wave travel. This
development of a proximal pressure trough has also been observed in a FLIP
prototype where there was an additional pressure sensor near the proximal end.
Due to inaccuracy in the sensor, the readings have not been plotted.
Qualitatively speaking, however, the readings showed a drop in pressure which
is supported by our calculations.

The work curves shown in the right column of
Fig.~\ref{fig:two_swallows_work_pressure} are computed from the estimated
fluid pressure and velocity fields. The curves show the same behavior observed
in the curves derived from the 1D model i.e. the work done to change the
kinetic energy of the fluid is minimal and the active work done is mostly lost
via viscous dissipation. The calculation of active and passive work done from
the 1D model in Section \ref{workcurves1dtheory} depends on the value of
$\theta$ which is completely known. However, the activation strength is not
available from real-world patient data, as such estimating the breakdown of
the pressure work term into an active and passive component is difficult. To
get around this hurdle, we use the FLIP data to get the tube's reference area
$(A_0)$ and the tube stiffness constant ($K$) when the esophagus is fully
relaxed. Once this information is obtained, the passive power was estimated as
\begin{equation}
\mathcal{P}_{pass} = \int_{0}^{L}p_{\text{pass}}\frac{\partial A}{\partial t}\mathrm{d}x = 
\int_{0}^{L}K\left(\frac{A}{A_{0}}-1\right)\frac{\partial A}{\partial t}\mathrm{d}x.
\end{equation}
The active work is then estimated by subtracting the passive work from
the total fluid pressure work. These estimated values of passive and active
work are then plotted in the work curves shown in Fig.~\ref{fig:sw1wrk}. The passive work is again seen to be
smaller than the magnitudes of active and viscous work. One observation
from these curves is that the total magnitude of work done for a
single healthy secondary peristaltic contraction is of the order of $10$
millijoules (mJ). Studies conducted using traction force sensors have been
able to quantify the propulsive force generated by the esophagus during
peristalsis \cite{Pouderoux1997}. On the lower end of the spectrum, the
esophagus was able to generate 0.11 N of force. During FLIP examinations,
around 12 cm of the bag lies within the segment of the esophagus that
undergoes peristaltic contraction. Using these values of force and distance, a
rough estimate of the work done was found to be 13.2 mJ which is of the same
order of magnitude as predicted by our model. {In a separate clinical
study, we compute the work done for several controls and patients belonging to
four disease groups to understand differences in peristaltic work done. \cite{AcharyaEsoWork2020}}

\section{Model Limitations}

{In this section, we discuss a few limitations of the 1D model developed to
study the FLIP device and its response to peristaltic contraction.}

{During particularly strong peristaltic activity, circular muscle
contraction can obliterate the esophageal lumen and completely cut off the
proximal and distal segments of the tube from each other. In this scenario,
both velocity $U$ and area $\alpha$ in the contraction region are zero and the
model becomes singular. However, in a majority of peristaltic contractions
studied with the FLIP, flow is observed to occur between chambers indicating
nonzero $\alpha$ and $U$ in this region.}

{Another avenue of exploration is the effect of nonlinear tube laws on the
quantity of work done during peristalsis. The complex fiber architecture of
the esophageal wall is one reason for this nonlinear behavior. The other
reason is the polyurethane bag used in the device. \textcolor{REDCOLOR2}{For bag volumes beyond 40
mL, there is greater stretching of the bag walls. When fully stretched, they
sharply resist further dilation.} This sudden increase in stiffness might have
a significant effect on the estimate of work done. The esophagus' material
properties vary along its length as well. The exact variation of these
properties is unknown, so as a first step, we assumed a simple linear tube law
to estimate work. Additionally, tube laws that incorporate viscoelasticity
will also significantly increase the amount of work done as the energy lost
within the walls due to hysteresis will be considered.} 

{Finally, the peristaltic activation wave does not have a constant
velocity. The wave can speed up or even completely stop depending on the
amount of obstruction it senses as a form of feedback regulation. In our
configuration, this can become quite important as the ends are closed and
continuous pumping will lead to a larger obstruction which can alter the speed
and intensity of contraction in subjects, directly affecting the work done by
the wave. Another detail that has not been considered is the precise
positioning of the bag within the esophagus. When deployed by the physician,
part of the bag straddles the Esophagogastric junction (EGJ) and the tip lies
within the stomach. The EGJ is a physiologically complex region that behaves
quite differently compared to the rest of the esophagus. In our work, we have
assumed that the entire bag rests within the esophagus.}


\section{Concluding remarks} 

In this work, we have analyzed a hitherto unstudied configuration of fluid
flow in an elastic tube. A simple reduced-order model is presented that is
able to predict fluid flow and the resultant tube wall deformation when a
peristaltic wave passes over a closed cylindrical tube. The model has been
compared to detailed three-dimensional immersed boundary simulations and is
shown to have satisfactory agreement with them. The system's response under
this set of operating conditions has been thoroughly quantified and visualized
as a regime map.

Finally, the system's utility is demonstrated by applying it to enhance the
data collected with balloon dilation catheters. Paired with this tool, the
device can now be thought of as having multiple pressure and velocity sensors
housed on the catheter. With the help of these quantities, an appropriate
mathematical foundation was laid to quantify the peristaltic work done. This
step has led us to estimate the work done by the esophagus during a healthy
contraction under these operating conditions. The device can now be used to
find work done in other peristaltic waves and comment on their capability to
propel fluid.

{The analysis has a few limitations due to the inherent complexity of the
combined organ-device system. However, the model presented in this work
provides a foundation on which future studies can be based and used to further
our understanding of esophageal pumping processes.}

\section*{Acknowledgment} 

This research was supported in part through the computational resources and
staff contributions provided for the Quest high performance computing facility
at Northwestern University which is jointly supported by the Office of the
Provost, the Office for Research, and Northwestern University Information
Technology.

This work also used the Extreme Science and Engineering Discovery Environment
(XSEDE) cluster Comet, at the San Diego Supercomputer Center (SDSC) through
allocation TG-ASC170023, which is supported by National Science Foundation
grant number ACI-1548562 \cite{xsede}.

\section*{Funding Data}
\begin{itemize}
\item National Institutes of Health (NIDDK grants R01 DK079902 and P01 DK117824; Funder ID: 10.13039/100000062)
\item National Science Foundation (OAC grants 1450374 and 1931372; Funder ID: 10.13039/100000105)
\end{itemize}

\section*{Disclosures}

Dustin A. Carlson, Peter J. Kahrilas, and John E. Pandolfino hold shared
intellectual property rights and ownership surrounding FLIP panometry systems,
methods, and apparatus with Medtronic Inc. No conflict of interest exists for
Shashank Acharya, Sourav Halder, Wenjun Kou and Neelesh A. Patankar.

\bibliographystyle{asmejour}   

\bibliography{pap_refs_asmejour} 

\end{document}